\documentclass{emulateapj}
\usepackage{apjfonts}
\usepackage{graphicx}
\usepackage{psfig}
\usepackage{amssymb}
\newcommand{\begit}{\begin{itemize}}
\newcommand{\enit}{\end{itemize}}
\newcommand{\begen}{\begin{enumerate}}
\newcommand{\enen}{\end{enumerate}}

\newcommand       \be           {\begin{equation}}
\newcommand       \ee           {\end{equation}}
\newcommand       \bea          {\begin{eqnarray}}
\newcommand       \eea          {\end{eqnarray}}

\newcommand       \cm		{\,{\rm cm }}
\newcommand       \km		{\,{\rm km }}

\newcommand       \yr		{\,{\rm yr }}
\newcommand       \mins		{\,{\rm min }}
\newcommand       \dday		{\,{\rm d }}

\newcommand       \hr		{\,{\rm hr }}

\newcommand       \La		{\,{\cal L }}
\newcommand       \G		{\,{\cal G }}
\newcommand       \Ha		{\,{\cal H }}
\newcommand       \kpc		{\,{\rm kpc }}

\newcommand       \ergs		{\,{\rm erg \,\, s}^{-1}}

\newcommand       \ic           {\imath_{crit}}
\newcommand{\beqa}{\begin{eqnarray}} 
\newcommand{\eeqa}{\end{eqnarray}}

\shorttitle{ON THE DYNAMICS OF UCXBs}
\shortauthors{Prodan \& Murray}

\begin{document}

\title{ON THE DYNAMICS OF ULTRA COMPACT X-RAY BINARIES:  4U 1850-087, 4U 0513-40 AND M15 X-2} 
\author{Snezana Prodan\altaffilmark{1} \& Norman Murray\altaffilmark{1,2}}

\altaffiltext{1}{Canadian Institute for Theoretical Astrophysics, 60
St.~George Street, University of Toronto, Toronto, ON M5S 3H8, Canada;
sprodan@cita.utoronto.ca} 
\altaffiltext{2}{Canada Research Chair in Astrophysics}

\begin{abstract}

In this work we extend our dynamical study of Ultra Compact X-ray Binaries (UCXB) 4U 1820-30 from  \citet{2012ApJ...747....4P} to three more
UCXBs in globular clusters: 4U 1850-087, 4U 0513-40 and M15 X-2. These
three UCXBs have orbital periods $\lesssim 20 \mins$. Two of them,  4U
1850-087 and 4U 0513-40, have suspected luminosity variations of order of $\sim
1\yr$. There is insufficient observational data to
make any statements regarding the long periodicity in the light curve of M15 X-2 at
this point. The properties of these three systems are quite similar to 4U
1820-30, which prompt us to model their dynamics in the same manner. As in
the case of 4U 1820-30, we interpret the suspected long periods as the
period of small oscillations around a stable fixed point in the Kozai
resonance.  We provide a lower limit on the tidal dissipation factor
$Q$ which is in agreement with results obtained for the case of 4U 1820-30.

\end{abstract}

\keywords{binaries: close --- stars: individual 4U1820-30, 4U 1850-087, 4U 0513-40, M15 X-2--- stars: kinematics and dynamics--- celestial mechanics}

\section{INTRODUCTION}

The possible existence of  long period ($\sim 100 \dday$) variations in the luminosity of UCXBs with orbital periods  $\lesssim 30\mins$ raises the possibility that the binary is orbited by a third body. The ratio between the orbital period and the period of the luminosity variations is too large to be due to any kind of accretion disc precession or change in the viewing angle \citep{1998MNRAS.299L..32L, 1999MNRAS.308..207W}. These long period variations in luminosity may be due to the actual change in the mass transfer rate of the binary. The presence of the third body may induce periodic oscillations in the eccentricity of the inner binary, which in turn will cause variations of the mass transfer rate, with the same period.  Following  \citet{2012ApJ...747....4P}, we show that this long period can be explained as the period of libration around the stable fixed point deep in a Kozai resonance. Since the expansion timescale of the inner binary is on the order of that of the accretion ($\sim 10^7\yr$), the action is indeed an adiabatic invariant, and as we demonstrated in detail in  \citet{2012ApJ...747....4P} the resonant trapping in libration around the fixed point in a Kozai resonance is a natural consequence. In contrast, tidal dissipation shrinks the semimajor axis. If tidal dissipation plays a primary role in how binary separation evolves, resonant trapping is no longer possible.

\subsection {4U 1850-087}

4U 1850-087 is a UCXB located in the galactic globular cluster NGC 6712.  The distance to the cluster is $6.8\kpc$ \citep{1975AJ.....80..427P, 1996AJ....112.1487H}. It was  first detected as an X-ray burster by \citet{1976IAUC.3010....1S}, which immediately indicates that the primary is a neutron star. The cluster centre and the source are separated by $~6''\pm6"$ \citep[$0.1\pm0.1$ core radii;][] {1983ApJ...275..105H}. This system has been observed since the very beginnings of observational X-ray astronomy, with many detections. It exhibits order of magnitude flux variations\citep{1978ApJS...38..357F, 1980ApJ...240L..27H, 1981MNRAS.197..865W, 1984ApJ...280..661P, 1984ApJS...56..507W, 1988MNRAS.232..551W, 1992ApJ...391..220K, 1997ApJS..109..177C, 2001ApJ...560L..59J}. \citet{1996MNRAS.282L..37H} reported a low amplitude periodicity the ultraviolet counterpart of this source, where their periodogram can be equally well fitted with a sinusoidal modulation at either of the two suspected orbital periods of $20.6\min$ or $13.2\mins$. However, as \citet{1996MNRAS.282L..37H} point out, a period as short as $13.2\mins$ is very close to the $11\mins$ period observed in 4U 1820-30 which is a much more luminous source. Since the larger luminosity corresponds to higher mass transfer rate, 4U 1850-087 would be underluminouns by a factor of $\sim 100$ for a $13.2\mins$ binary  \citep{1996MNRAS.282L..37H,1987ApJ...322..842R}. Hence, $20.6\mins$ is interpreted as an orbital period even though it has yet to be confirmed.  Such a short period implies that a mass losing companion has to be a degenerate and low mass star. Following the consideration of 4U 1820-30 from \citet{1987ApJ...322..842R}, \citet{1996MNRAS.282L..37H} derive the mass and  the radius of the secondary to be $0.04 M_{\odot}$ and $0.04 R_{\odot}$ respectively under the assumption that the secondary is a fully degenerate helium white dwarf. Additionally, assuming a low-mass white dwarf donor and mass transfer driven by gravitational radiation, \citet{1996MNRAS.282L..37H}  showed that the X-ray luminosity of this system  is that expected for period of $20.6\mins$.

An interesting feature of 4U 1850-087 is a possible long period of $0.72\yr$ reported by \citet{1984ApJ...280..661P} at which luminosity varies by a factor of $2-3$. Even though, as we show further in the paper, this long period fits well in the dynamical picture given by our model, it needs to be further verified observationally.

\subsection{4U 0513-40}

4U 0513-40 is a low mass X-ray binary in the globular cluster NGC 1851. The distance to the source is $12\kpc$ \citep{1996AJ....112.1487H}. Far ultraviolet photometry obtained by HST revealed a $17\mins$ orbital modulation
\citep{2009ApJ...699.1113Z}. Observations with \emph{BeppoSAX}, \emph{Chandra}, \emph{XMM-Newton}, \emph{INTEGRAL} confirmed the $17\mins$ periodic sinusoidal signal in soft X-ray \citep{2011MNRAS.414L..41F}. The system is known to be an X-ray burster  \citep{2008ApJS..179..360G, 2001ApJ...550L.155H, 2011MNRAS.414L..41F} which indicates that the primary is a neutron star. The short orbital period suggests a low mass white dwarf secondary of $\sim 0.05 M_{\odot}$, as implied from mass- radius relation for $17\min$ period binaries from \citet{2003ApJ...598.1217D}.

4U 0513-40 shows very interesting variability in the X-ray flux on two different time scales \citep{2010MNRAS.406.2087M}; a factor of $\sim 10$ variation on timescales of weeks and  the variation of a factor of $\sim 2$ in the luminosity when averaged over $~1\yr$ \citep{2010MNRAS.406.2087M}. This long time scale variation points toward a modulation in the mass transfer rate, even though the mean luminosity agrees with that predicted by  gravitation radiation driven evolutionary scenario.

\subsection{M15 X-2}

M15, at a distance of $10.4\kpc$ \citep{1996AJ....112.1487H}, is the only globular cluster associated with our galaxy  known to house two bright LMBXs. In the early X-ray studies, a single source 4U 2127+119 was first identified with the optical counterpart AC 211 by \citet{1984A&A...138..415A} and further confirmed by a spectroscopic study by \citet{1986Natur.323..417C} showing signatures of an LMXB. A modulation in the optical and the X-ray flux revealed the orbital period of $17.1\hr$ \citep[and references within]{1993A&A...270..139I}. AC 211 is among the brightest LMBXs in the optical and at the same time it has a low X-ray luminosity $\sim 10^{36} \ergs$; the high optical-to -X-ray luminosity ratio implied that a very luminous central X-ray source is hidden behind the accretion disk \citep{1984A&A...138..415A}. However, when X-ray bursts were detected by \emph{Ginga} satellite \citep{1990Natur.347..534D} and later on with \emph{Rossi X-ray Timing Explorer} \citep{2001ApJ...562..957S}, this conclusion was highly dubious. This puzzling behaviour was finally understood when \emph{Chandra} observations resolved 4U 2127-119 into \emph{two} X-ray sources \citep{2001ApJ...561L.101W}. One source is of course already known LMBX AC 211, while the second one, named M15 X-2 which is the one producing X-ray bursts, is 2.5 times brighter in X-rays than AC 211. The source is located $3\farcs4$ from the center of M15. The optical and the FUV counterparts of M15 X-2 were identified in HST data by \citet{2001ApJ...561L.101W} and \citet{2005ApJ...634L.105D} respectively resulting in a determination of the orbital period of $22.6\min$ by \citet{2005ApJ...634L.105D}. The donor corresponding to such a short period is a white dwarf of mass $0.02 M_{\odot} \leq M_{2, min} \leq 0.03 M_{\odot}$ and of radius $0.02 R_{\odot} \leq R_{2, min} \leq 0.03 R_{\odot}$. The existence of the X-ray bursts is consistent with neutron star primary \citep{2005ApJ...634L.105D}.

So far no long period luminosity variations have been reported in M15 X-2. In section \ref{sec:numerics} we discuss a constraint on the shortest expected long period assuming the conservative mass transfer given by our model.

\subsection{ Plan of the paper}

In this paper we apply the dynamical model described in  \citet{2012ApJ...747....4P} on the three known UCXBs in the globular clusters just described: 4U 1850-087, 4U 0513-40 and M15 X-2. We demonstrate that the suspected long period of 4U 1850-087 and 4U 0513-40 can be explained as libration in Kozai resonance with the period of small oscillations around the fixed point. Our model gives a prediction for a yet undetected long period of M15 X-2. As shown in  \citet{2012ApJ...747....4P} the dynamical history of these systems is a consequence of the interplay of two effects. The mass transfer via Roche lobe overflow drives the systems into a resonance, tidal dissipation tends to damp the mutual inclination close to the Kozai critical inclination. The interplay between these two effects allows us to infer a constraint on the tidal dissipation factor Q for white dwarf donors in these systems. In section \ref{sec:review} we give a review of the dynamical model developed in  \citet{2012ApJ...747....4P} and estimates for the systems' parameters.  The numerical results are presented in section \ref{sec:numerics}. In section \ref{sec:Qconstraint} we constrain the ratio of the tidal dissipation factor  and the tidal Love number, $Q/k_2$ of the white dwarf donors. We end with a discussion in section \ref{sec:discussion}.

\section{OVERVIEW OF OUR DYNAMICAL MODEL}\label{sec:review}

In the work of \citet{2012ApJ...747....4P} on the dynamics of 4U 1820-30, we argue that the observed long period modulation of the luminosity ($\sim 170$ day) is caused by the presence of a third body orbiting the center of mass of the binary. Variations in the eccentricity of the inner binary are associated with libration around the stable fixed point deep in the Kozai resonance.  Kozai resonance is $1 : 1$ resonance between the precession rate of the longitude of periastron $\dot \varpi$ and the precession rate of the longitude of the ascending node $\dot \Omega$ of the inner binary. The condition for Kozai resonance is satisfied only in cases where the mutual inclination is above its critical value of $39^\circ.2$.  Taking into account the presence of additional precessions, we demonstrate that the luminosity modulation arises from the these eccentricity variations. The additional precessions are due to tidal and rotation distortion of the secondary, tidal dissipation and apsidal precession due to general relativistic effects (GR). Here we list the equations for all four precession rates:

\bea \label{eqn: eqn for omega_dots}
\dot{\omega}_{Kozai}&=&{3\over4}n
\left({m_3\over m_1+m_2}\right)
\left({a\over a_{out}}\right)^3
{1\over(1-e_{out}^2)^{3/2}}\times
{1\over\sqrt{1-e^2}}\nonumber\\
&&\qquad\left[ 2(1-e^2)+5\sin^2\omega(e^2-\sin^{2}i)\right] \label{eq:kozai}\\
\dot{\omega}_{GR}&=&{3\over2}n\left({m_1+m_2\over m_1}\right)
\left({r_s\over a}\right)
{1\over(1-e^2)}\label{eq:GR}\\
\dot{\omega}_{TB}&=&{15\over16}n\,k_2{m_1\over m_2}
\left({R_2\over a}\right)^5{8+12e^2+e^4\over(1-e^2)^5}\label{eq:tb}\\
\dot{\omega}_{RB}&=&{n\,k_2\over4}{m_1+m_2\over m_2}
\left({R_2\over a}\right)^5{1\over(1-e^2)^2}\nonumber\\
\times&\Bigg[&
\left(2\tilde\Omega^{2}_{h}-\tilde\Omega^{2}_{e}-\tilde\Omega^{2}_{q}\right)+2\tilde\Omega_{h}\cot i\left(\tilde\Omega_{e}\sin\omega +\tilde\Omega_{q}\cos\omega\right)
\Bigg].\label{eq:rb}
\eea
Masses of the primary, secondary and tertiary are $ m_1$, $m_2$ and $m_3$. The orbital elements of the inner binary are following: eccentricity $e$, semimajor axis $a$ , mutual inclination between the inner binary and the outer binary orbit $i$, the argument of periastron $\omega$, the longitude of ascending node $\Omega$. G is Newtons constant and c is the speed of light. $n = 2\pi/P = [GM/a^3]^{1/2}$ denotes the mean motion of the inner binary. $k_2$ is the tidal Love number and $R_2$ is the radius of the white dwarf.The quantity $r_s = 2Gm_1/c^2$ in equation \ref{eq:GR} is the Schwarzschild radius of the neutron star.  $a_{out}$ and $e_{out}$  are semimajor axis and eccentricity of the outer binary. $\tilde\Omega_e$, $\tilde\Omega_h$ and $\tilde\Omega_h$ are spin projections onto the triad defined by Laplace-Runge vector $e$, the total angular momentum vector $h$ and their cross product $q=e\times h$. 

The sign of the Kozai term depends on $\sin i$, while the contribution from the tidal bulge of the secondary and the GR term are always positive. The contribution from the rotational bulge of the secondary is positive if the secondary is tidally locked and if the spins are aligned; in the opposite case it is negative. Throughout the paper we adopt $k_2=0.01$ for helium white dwarf.  Such a value of $k_2$ is obtained for the helium white dwarf in 4U 1820-30, assuming that it is a fluid objects, as the ratio of the potential due to the perturbed mass distribution, to the external potential causing the perturbed mass (Arras, private communication). Assuming that the system is tidally locked, to produce these observed long period modulations, we are looking toward the cancellation better than $10\%$ between the Kozai term and all other terms. It turns out that the dominant contribution, other than the Kozai precession, comes from the tidal bulge for our fiducial value of $k_2$. Fine tuning of inclination, required in such situation, can be avoided if the system librates around the stable fixed point. In this case  precession rates sum to zero. Therefore, these long periods of libration are associated with the period of small oscillations around the fixed point. 

The maximum allowed eccentricities for the system in consideration listed in Table 1, are of order of $\sim 0.05$. These values are given by \citet{2005MNRAS.358..544R}  assuming the conservative mass transfer and lack of mass loss via the  $L_2$ Lagrangian point at the mass ratios of these three systems. \citet{2012ApJ...747....4P} demonstrate that the periods of small oscillation of order of several hundred days require such small eccentricities. When the mutual inclination is close to the Kozai critical value such small eccentricity oscillations arise naturally.  

As shown in  \citet{2012ApJ...747....4P}, via both analytical and numerical calculations, in order to trap the system in the resonance and then to maintain it trapped for at least $10^5\yr$ (which is about $\sim 1\%$ of the lifetime for such systems), the semimajor axis has to expand. In another words the mass transfer has to dominate the evolution of the semimajor axis. Since the semimajor axis expands on a timescale much larger than any orbital or precession timescale in the system, the action of the Hamiltonian that describes the dynamics of such systems is an adiabatic invariant. For detailed discussion on the adiabatic invariant we refer reader to \citet{2012ApJ...747....4P}. The semimajor axis expansion of the inner binary lead to the increase of the torque between the two orbits which is equivalent to deepening of the Kozai potential. The fact that the action is an adiabatic invariant means that the action of all orbits other than the separatrix remain constant, while the action of the separatrix increases. As the action of the separatrix grows, eventually it will exceed the action of the initially circulation orbit. At that point the orbit in question is captured in the resonance and starts to librate.The semimajor axis expansion eventually leads the orbit close to the fixed point where is librates with the period of small oscillations. 
 
 On the other hand, if tidal dissipation, which has a tendency to shrink the semimajor axis, plays dominant role in the evolution of the semimajor axis, the system does not remain captured in the resonance. Intense tidal dissipation reduces the eccentricity of the inner pairwithin the $10^{-3}$ of the system's lifetime leading toward the loss of long period modulation of luminosity. Therefore, we would have to be extremely fortunate to observe the system during such a short phase when the eccentricity modulation induced by the tertiary is important. Hence we assert that the effect causing the trapping of the system deep in Kozai resonance is the expansion of the semimajor axis driven by Roche lobe overflow. While in the resonance, angular momentum is transferred back and forth periodically between the inner binary and the third star. Such exchange of angular momentum does not affect by any means the semimajor axis of both binaries. However, when the forced eccentricity is at its maximum and the mutual inclination at its minimum, strong tidal dissipation reaches its maximum in removing the energy from the inner orbit. Such coupled Kozai-tidal evolution brings the mutual inclination toward its critical value \citep[$\sim 40^\circ$,] []{2007ApJ...670..820W, 2007ApJ...669.1298F}.

For additional discussion on Kozai mechanism in the presence of external forces and its application to different systems we refer reader to: \citet{2001ApJ...562.1012E}, \citet{2002ApJ...578..775B}, \citet{2002ApJ...576..894M}, \citet{2003ApJ...589..605W}, \citet{2007ApJ...670..820W}, \citet{2007ApJ...669.1298F},  \citet{2010ApJ...713...90A}, \citet{2011Natur.473..187N}, \citet{2012ApJ...757...27A}, \citet{2012arXiv1206.4316N}, \citet{KatzDong2012}, \citet{2013MNRAS.431.2155N}, \citet{Prodan2013}, \citet{2013MNRAS.430.2262H}, \citet{2013ApJ...766...64S}, \citet{2014ApJ...781...45A}, \citet{2014arXiv1405.6029P}.

\subsection{Estimating the mass, the radius and the mass transfer rate of the white dwarf secondary }\label{sec:masses}
To model the evolution of the system due to the mass transfer we follow the prescription of \citet{1987ApJ...322..842R}. Assuming that the secondary is a polytrope with index $n=3/2$, completely degenerate and hydrogen depleted, its mass-radius relation is given by \citep{1987ApJ...322..842R}:
\be\label{eqn:mass-radius}
  \frac{R_2}{R_{\odot}}=0.0128\left(\frac{m_2}{M_{\odot}}\right)^{-\frac{1}{3}}
\ee
Since the secondary fills its Roche lobe, we have:
\be\label{eqn:fill_roche}
  P=\frac{9\pi}{\sqrt2}(Gm_2)^{-\frac{1}{2}}R^{\frac{3}{2}}_2
\ee  
Combining equation \ref{eqn:mass-radius} and equation \ref{eqn:fill_roche} we obtain the mass-period relation \citep{1987ApJ...322..842R}:
\be\label{eqn:mass-period}
  m_2=0.769\left(\frac{P}{min}\right)^{-1} M_{\odot}
\ee
Equation \ref{eqn:mass-period} puts a constraint on the mass of the donor star and knowing its mass we constrain the radius of the donor using equation \ref{eqn:mass-radius}. For each binary we list the mass and the radius of the donor star, as well as X-ray luminosity, maximum eccentricity and the long period, in Table 1. The estimated masses and radii of the secondary for each system listed in Table 1 are in agreement with the approximate values quoted in the introduction for the masses of the secondary given by the mass-radius relation for appropriate orbital period by \citet{2003ApJ...598.1217D}. For the mass of the neutron star primary we adopt a canonical mass of $1.4M_{\odot}$.   


\begin{table*}
\begin{center}
\begin{tabular}{|l|l|l|l|}
\tableline
\multicolumn{4}{|c|}{TABLE 1. Constrained binary parameters} \\ \hline
\multicolumn{4}{|c|}{4U 1850-087} \\
\tableline
\tableline
Symbol & Definition & Value & Citation \\ \hline
$m_2$ & White dwarf (secondary) mass & $0.04 M_{\odot}$ & \citet{2001ApJ...550L.155H}\\
$R_2$ & White dwarf (secondary) radius & $2.78\times10^9\cm$ &\\
$L_X$ & X-ray luminosity & $1\times10^{36}\ergs$& \citet{1992ApJ...391..220K} \\
$e_{max}$ & Maximum inner binary eccentricity& $0.05$ & \citet{2005MNRAS.358..544R}\\
$P_0$ & Long period & $0.72\yr$ & Priedhorsky (1986) \\ \hline
\multicolumn{4}{|c|}{4U 0513-40} \\
\tableline
\tableline
Symbol & Definition & Value & Citation \\ \hline
$m_2$ & White dwarf (secondary) mass & $0.045 M_{\odot}$ &\\
$R_2$ & White dwarf (secondary) radius & $2.5\times10^9\cm$ &\\
$L_X$ & X-ray luminosity & $3\times10^{36}\ergs$& \citet{1995MNRAS.273..201C} \\
$e_{max}$ & Maximum inner binary eccentricity& $0.05$ & \citet{2005MNRAS.358..544R}\\
$P_0$ & Long period & $1\yr$ &\citet{2010MNRAS.406.2087M} \\ \hline
\multicolumn{4}{|c|}{M15 X-2} \\
\tableline
\tableline
Symbol & Definition & Value & Citation \\ \hline
$m_2$ & White dwarf (secondary) mass & $0.034 M_{\odot}$ &\\
$R_2$ & White dwarf (secondary) radius & $2.75\times10^9\cm$ &\\
$L_X$ & X-ray luminosity & $0.74\times10^{36}\ergs$& \citet{1996MNRAS.282L..37H} \\
$e_{max}$ & Maximum inner binary eccentricity& $0.04$ & \citet{2005MNRAS.358..544R}\\
$P_{0,min}$ & Minimum long period& $1\yr$ &\\ \hline
\tableline
\tableline
\end{tabular}
\end{center}
\end{table*}

\begin{figure}
\epsscale{0.8} 
\plotone{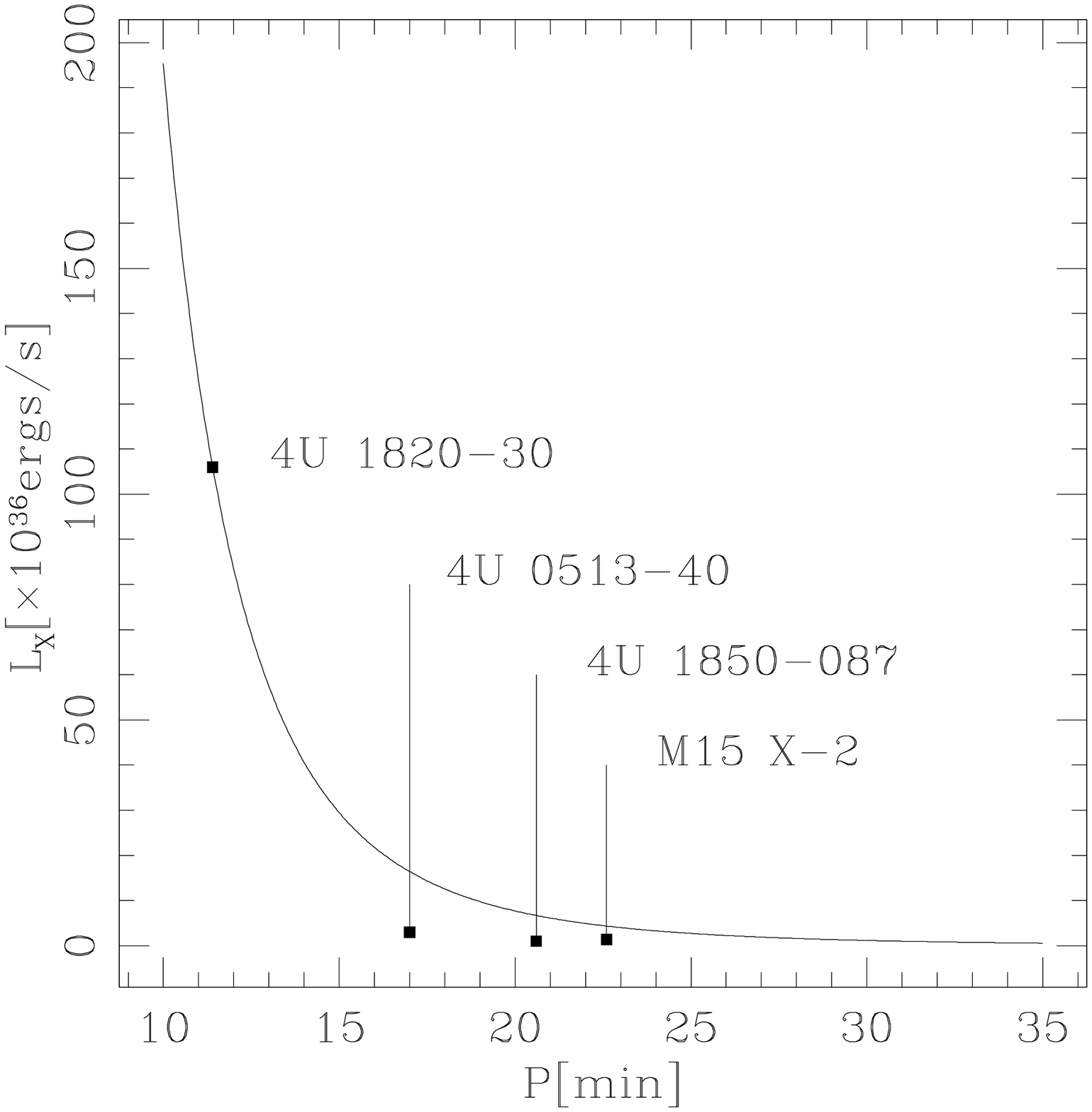}
\caption{The X-ray luminosity, $L_X$, versus the orbital period of the binary. The solid line represents the equation \ref{eqn:luminosity} while dots are averaged observed X-ray luminosities. As shown, equation \ref{eqn:luminosity} is not in contradiction with the X-ray luminosity of 4U 1850-087, 4U 0513-40 and M15 X-2.
 \label{Fig:LvsP}}
\end{figure}

The mass transfer rate is given by \citep[for detailed derivation see][] {1987ApJ...322..842R}:
\bea\label{eqn:mdot}
  \dot{m}_2&=&6.21\times10^{-4}\left(\frac{m_1}{M_{\odot}}\right)^{\frac{2}{3}}\left(\frac{P}{min}\right)^{-\frac{14}{3}}\frac{M_{\odot}}{\yr}\nonumber\\
  &=&1.23\times10^{-30}\left(\frac{AU}{a}\right)^{-7}\left(\frac{m_1}{M_{\odot}}\right)^3\frac{M_{\odot}}{\yr}
\eea  
Equation \ref{eqn:mdot} gives the following expression for X-ray luminosity \citep{1987ApJ...322..842R, 2001ApJ...550L.155H}:
\be\label{eqn:luminosity}
 L_X=1.06\times10^{38}\left(\frac{m_1}{1.4M_{\odot}}\right)^{\frac{5}{3}}\left(\frac{R_1}{10\km}\right)^{-1}\left(\frac{P}{11.4\min}\right)^{-\frac{14}{3}}\ergs
\ee
Both equation \ref{eqn:mdot} and equation \ref{eqn:luminosity} are scaled with respect to the parameters of 4U 1820-30. We plot equation \ref{eqn:luminosity} in Figure \ref{Fig:LvsP} and we demonstrate that this equation, originally derived to describe the evolution of 4U 1820-30, is not in contradiction with the X-ray luminosity of all three UCXBs: 4U 1850-087, 4U 0513-40 and M15 X-2.   

In order to account for small eccentricity of the inner orbit  when calculating $\dot{m}_2$ in our numerical model instead of semimajor axis $a$ we use periapse distance $a(1-e)$. Therefore equation \ref{eqn:mdot} becomes:

\be
  \dot{m}_2=1.23\times10^{-30}\left(\frac{AU}{a(1-e)}\right)^{-7}\left(\frac{m_1}{M_{\odot}}\right)^3\frac{M_{\odot}}{\yr}.
\ee  

\subsection{The eccentricity and the period of small oscillations of the inner binary}

We use Delaunay variables to characterize the motion of the inner pair: ,the argument of periastron $\omega$, the longitude of the ascending node
$\Omega$ and the mean anomaly $l$. In the Hamiltonian averaged over $l$ and $l_{out}$ only $\omega$ appears . Their respective conjugate momenta are:
\bea \La &=&m_1
m_2 \sqrt{\frac{Ga}{m_1+m_2}}\\ \G &=&\La\sqrt{1-e^2}\\ \Ha
&=&\G\cos{i}.  \eea
The averaged Hamiltonian is given by \citep{1997AJ....113.1915I, 2000ApJ...535..385F, 2007ApJ...669.1298F, 2012ApJ...747....4P}:
\bea \label{eqn:hamiltonian} 
H&=&\frac{-3A}{2}\Bigg[-{5\over3}-3\frac{\Ha^{2}}{\La^{2}}+\frac{\G^{2}}{\La^{2}}+5\frac{\Ha^{2}}{\G^{2}}+5\cos2\omega\left(1-\frac{\G^{2}}{\La^{2}}-\frac{\Ha^{2}}{\G^{2}}+\frac{\Ha^{2}}{\La^{2}}\right)\Bigg]\nonumber\\ &&-B\frac{\La}{\G}-k_2C\left(35\frac{\La^{9}}{\G^{9}}-30\frac{\La^{7}}{\G^{7}}+3\frac{\La^{5}}{\G^{5}}\right)
-k_2D\frac{\La^{3}}{\G^{3}}, 
\eea 
where the term proportional to $A$ is the Kozai term and the term
proportional to $B$ is GR apsidal precession. The terms proportional to $C$ and $D$
correspond to the tidal and rotational bulges. The expressions for the
constants are:
\bea 
A&=&{1\over8}\Phi\,
{m_2m_3\over(m_1+m_2)^2}
\left({a\over a_{out}}\right)^3
{1\over(1-e^2_{out})^{3/2}}\\ 
B&=&{3\over 2}\Phi
 {m_2\over m_1} {r_s\over a}\label{eq:GRa}\\
C&=&{1\over 16}\Phi
{m_1\over m_1+m_2}\left({R_2\over a}\right)^{5}\\
D&=&{1\over12}\Phi
\left({R_2\over a}\right)^5f(\tilde\Omega_{spin})\label{eq:spin},
\eea 
where
\be  
\Phi\equiv{G(m_1+m_2)m_1\over a}.
\ee  
Eccentricity and the semimajor axis of the outer body's
orbit are: $e_{out}$ and $a_{out}$. We denote the Schwarzschild
radius of the neutron star by $r_s\equiv2Gm_1/c^2$.
 
To obtain the the frequency or the period of the small oscillation, we evaluate the second derivative of the Hamiltonian given by equation \ref{eqn:hamiltonian} at the fixed point:
\bea \label{eqn:period} 
\omega_0=&\omega_A&\Bigg[\left(18+90{\Ha^2\La^2\over G_f^4}\right)+2{B\over A}\frac{\La^3}{G_f^3}\nonumber\\&&+k_2{C\over
    A}\left(3150\frac{\La^{11}}{G_f^{11}}-1680\frac{\La^9}{G_f^9}+90\frac{\La^7}{G_f^7}\right)\nonumber\\ &&+12k_2{D\over
    A}\frac{\La^5}{G_f^{5}}\Bigg]^{1/2}\times e_f\sin i_f,
\eea 
where $e_f$ and $\ic$ are the eccentricity of the fixed point and the critical mutual inclination given by \citep{2012ApJ...747....4P}:
\bea \label{eqn:efix}
e_f& =& \sqrt{30\left[\cos^2\ic-\cos^2i\right]\over{60\frac{\Ha^2}{\La^2}+{3\over2}{B\over A}+840k_2{C\over A}+\frac{15}{2}k_2{D\over A}}}\\
\cos^2\ic &=& {3\over 5}-{1\over 30}{B\over A} - 4k_2 {C\over A} -{1\over 10} k_2 {D\over A}.
\eea 
The argument of periastron of the fixed point is $\omega_f=90^\circ, 270^\circ$.

Under the assumption that there is no mass loss through the Lagrangian point $L_2$, we list in Table 1 the maximum possible eccentricities in the inner binaries given by \citet{2005MNRAS.358..544R} for the mass ratio of the binary. 4U 1850-087 and 4U 0513-40 have suspected but yet not confirmed long periods of $0.72\yr$ \citep{1986Ap&SS.126...89P} and $1\yr$ \citep{2010MNRAS.406.2087M} respectively. Their eccentricities are well below the maximum possible eccentricities given in Table 1. Adopting the listed value for the maximum eccentricity, we estimate using equations \ref{eqn:efix} and \ref{eqn:period} the value for the minimum possible period of small oscillations of M15 X-2 to be of order of $1\yr$. This estimate provides values on the verge of overflowing $L_2$ point, hence in reality we would expect this period to be longer and the eccentricity to be smaller.

\section{NUMERICAL RESULTS} \label{sec:numerics}

In our numerical calculation the gravitational impact due the third star is calculated in the quadrupole approximation that includes the Kozai resonance described before. We demonstrated in  \citet{2012ApJ...747....4P} that the results do not change qualitatively due to the introduction of the octupole approximation. We derive equations of motion from the Hamiltonian averaged over the orbital periods of the inner and the outer binary. The equations of motion, given in appendix A of \citet{2012ApJ...747....4P}, incorporate: 
\begin{itemize}
\item periastron advance due to general relativity;
\item periastron advance caused by quadrupole distortions of the white dwarf secondary due to tides and rotation;
\item orbital shrinkage of the inner binary orbit due to tidal dissipation in the white dwarf secondary;
\item orbital angular momentum loss due to gravitational radiation;
\item conservative mass transfer from the white dwarf secondary to the neutron
  star primary driven by the emission of gravitational radiation.
\end{itemize}

The initial conditions that give us appropriately long periods for each binary are listed in tables 2-4. These parameters are used throughout this chapter unless otherwise stated.  


\begin{table*}
\begin{center}
\begin{tabular}{|l|l|l|l|}
\tableline
\multicolumn{4}{|c|}{TABLE 2. 4U 1850-087: System parameters} \\
\tableline
\tableline
Symbol & Definition & Value & Citation \\ \hline
$m_1$ & Neutron star (primary) mass & $1.4 M_{\odot}$ &\\
$m_2$ & White dwarf (secondary) mass & $0.04 M_{\odot}$ &\citet{2001ApJ...550L.155H}\\
$m_3$ & Third companion mass & $0.55 M_{\odot}$& \\
$a_1$ & Inner binary semimajor axis & $1.95\times10^{10}\cm$ & \citet{1996MNRAS.282L..37H}\\
$a_{out}$ & Outer binary semimajor axis & $8.78a_1$ &\\
$e_{in, 0}$ & Inner binary initial eccentricity & $0.018$ &\\
$e_{out, 0}$ & Outer binary eccentricity & $10^{-4}$ &\\
$i_{init}$ & Initial mutual inclination & $44.657^{o}$  & \\
$\omega_{in, 0}$ & Initial argumet of periastron & $90^{o}$ &\\
$\Omega_{in}$ & Longitude of ascending node & $0$ &\\
$R_2$ & White dwarf radius & $2.78\times10^9 \cm$ &see section \ref{sec:masses} \\
$k_2$ & Tidal Love number & $0.01$ & Arras (private communication)\\
$Q$ & Tidal dissipation factor & $6\times10^7$ &\\
\tableline
\tableline
\end{tabular}
\end{center}
\end{table*}

\begin{figure}
\epsscale{0.7} 
\plotone{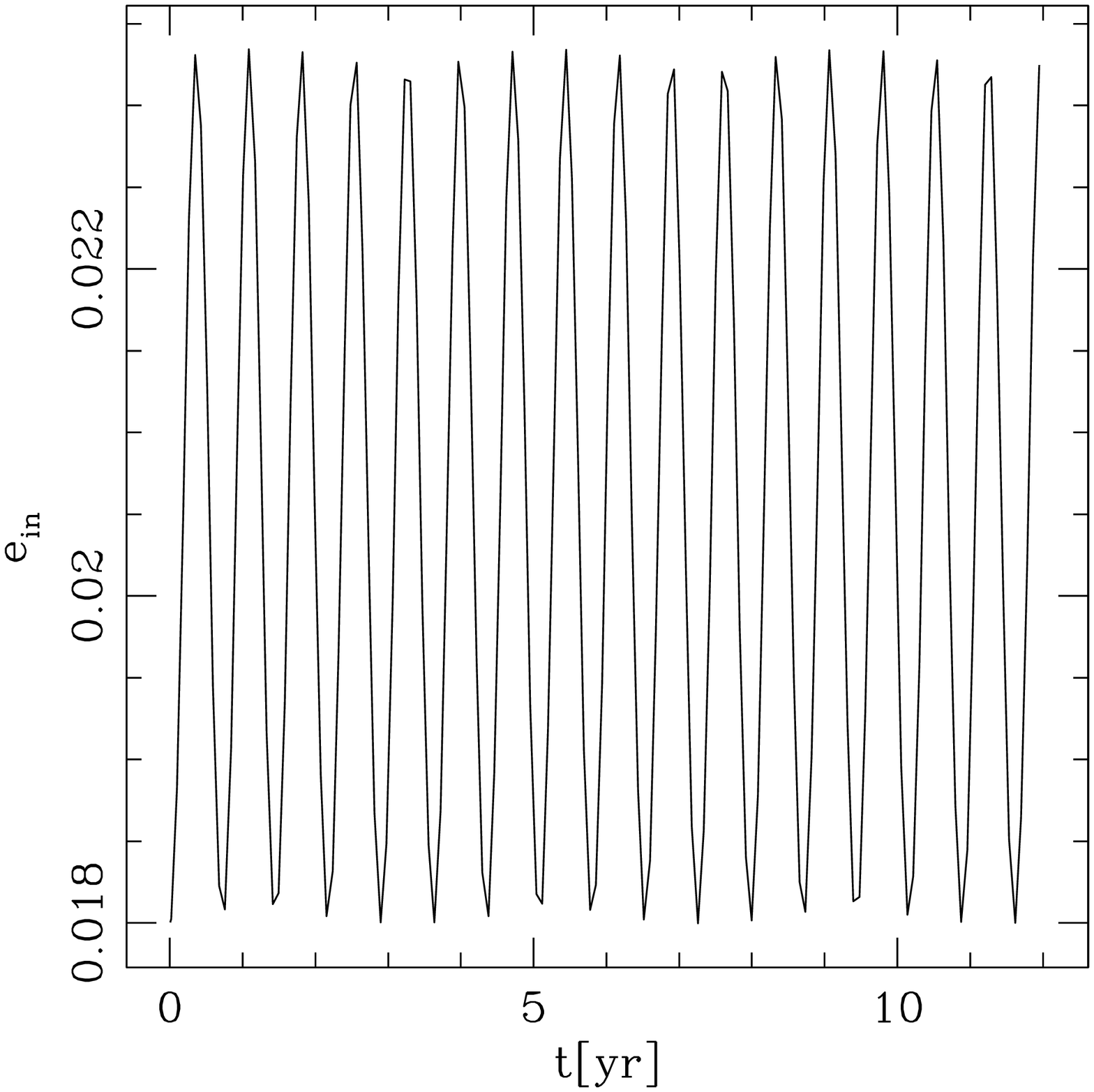}
\plotone{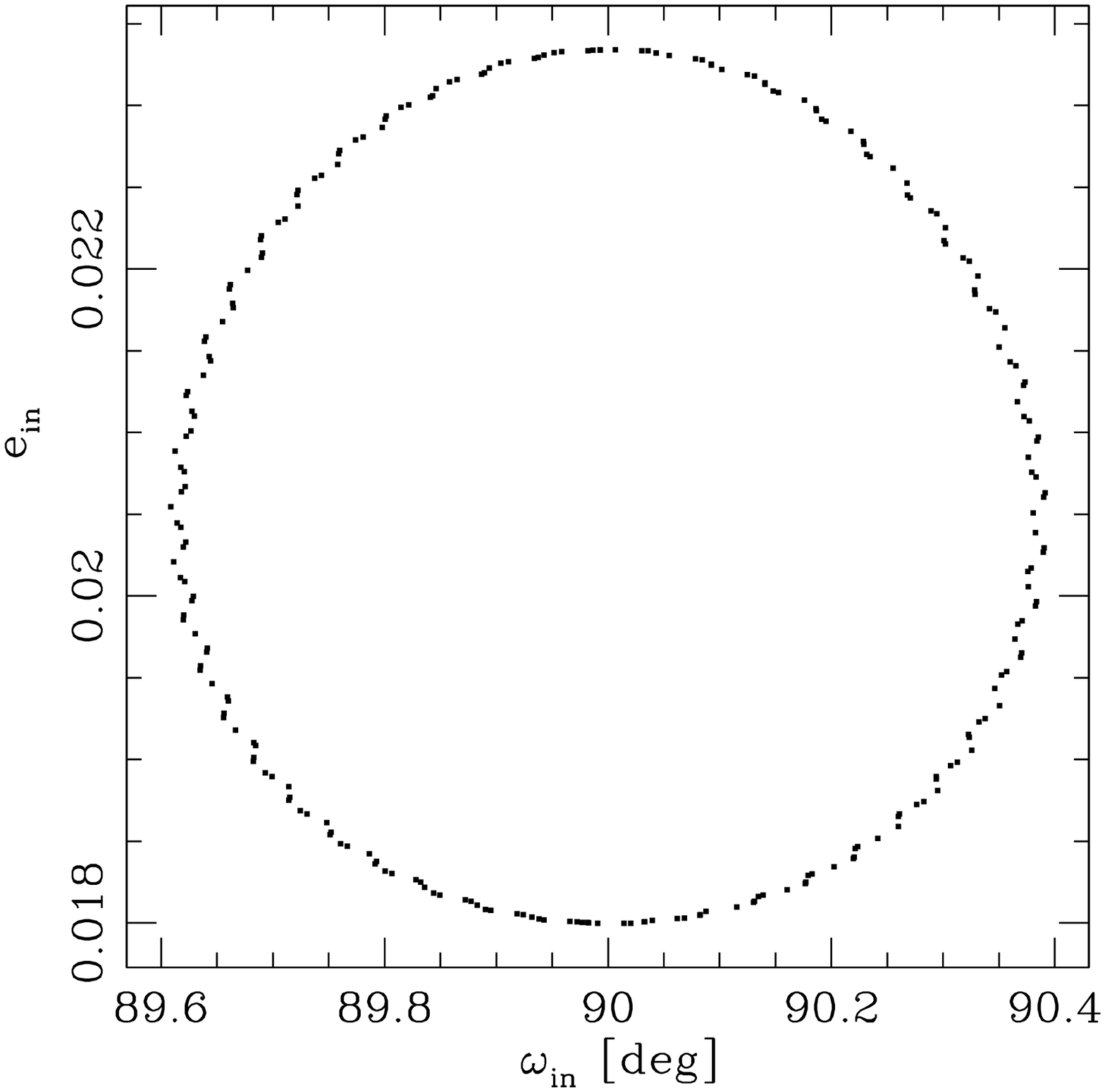}
\caption{4U1850-087:  Upper panel shows the eccentricity as a function of time while the lower panel shows the phase space, $e$ versus $\omega$, for the fiducial
  model of the system. The long period -- i.e. the period of the eccentricity oscillations -- is $0.72\yr$. The amplitude of the eccentricity oscillation is large enough to
  give rise to the observed factor of $2-3$ variation in luminosity.
 \label{Fig:4U18e}}
\end{figure}
\begin{figure}
\epsscale{0.7} 
\plotone{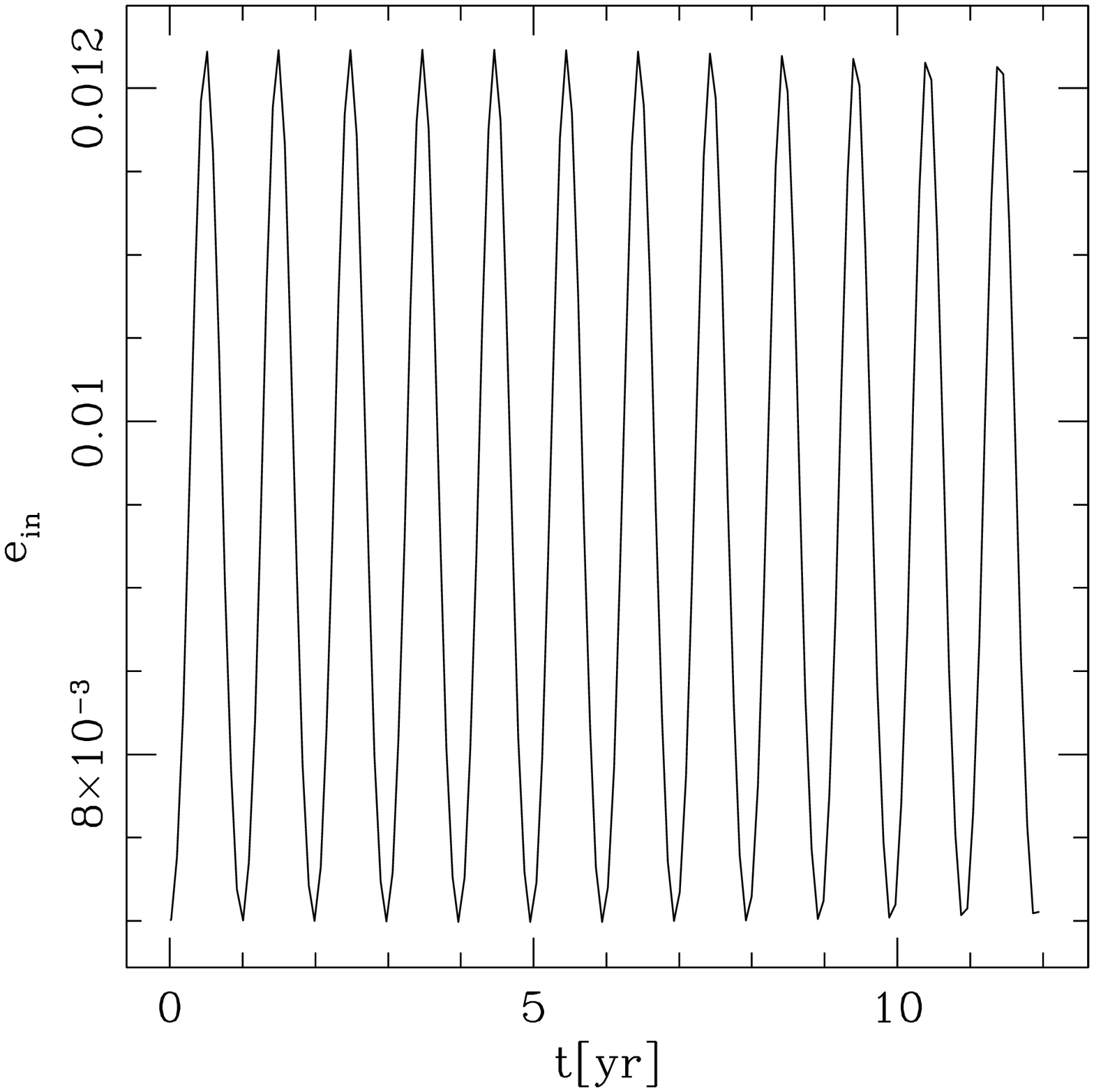}
\plotone{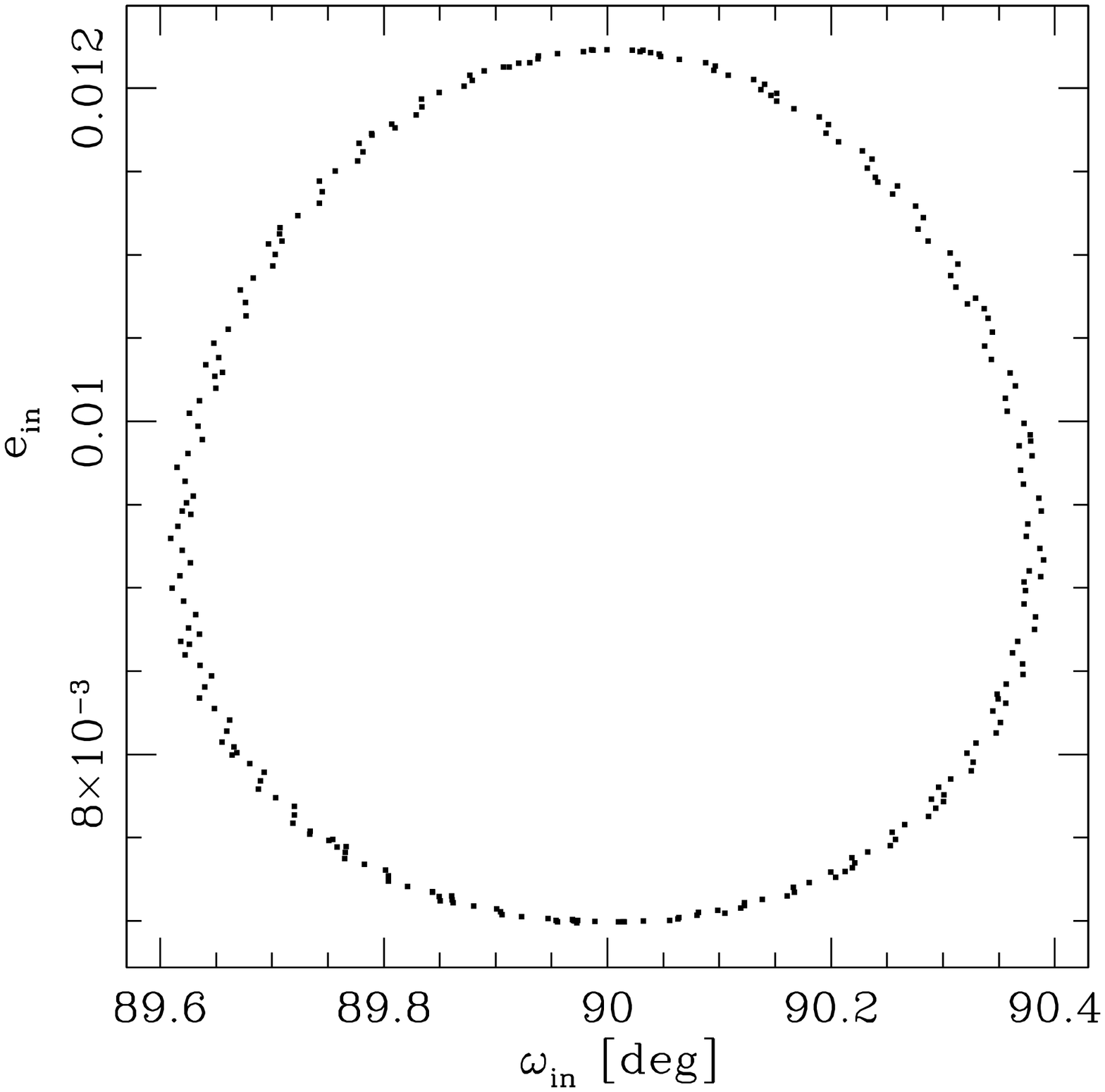}
\caption{4U 0513-40: Same as figure \ref{Fig:4U18e}.  The long period -- i.e. period of the eccentricity oscillations -- is $1\yr$. 
 \label{Fig:4U05e}}
\end{figure}
\begin{figure}
\epsscale{0.7} 
\plotone{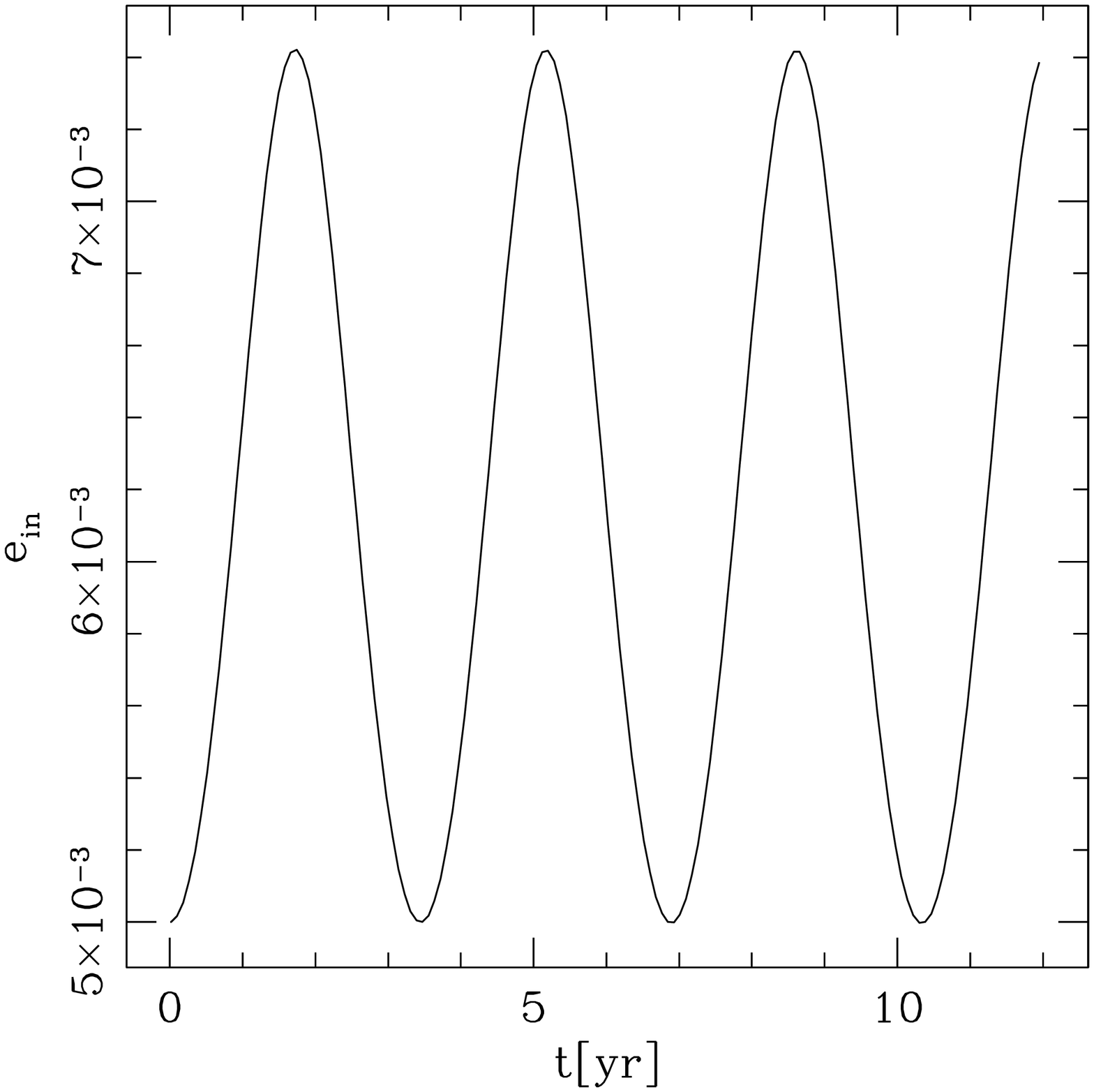}
\plotone{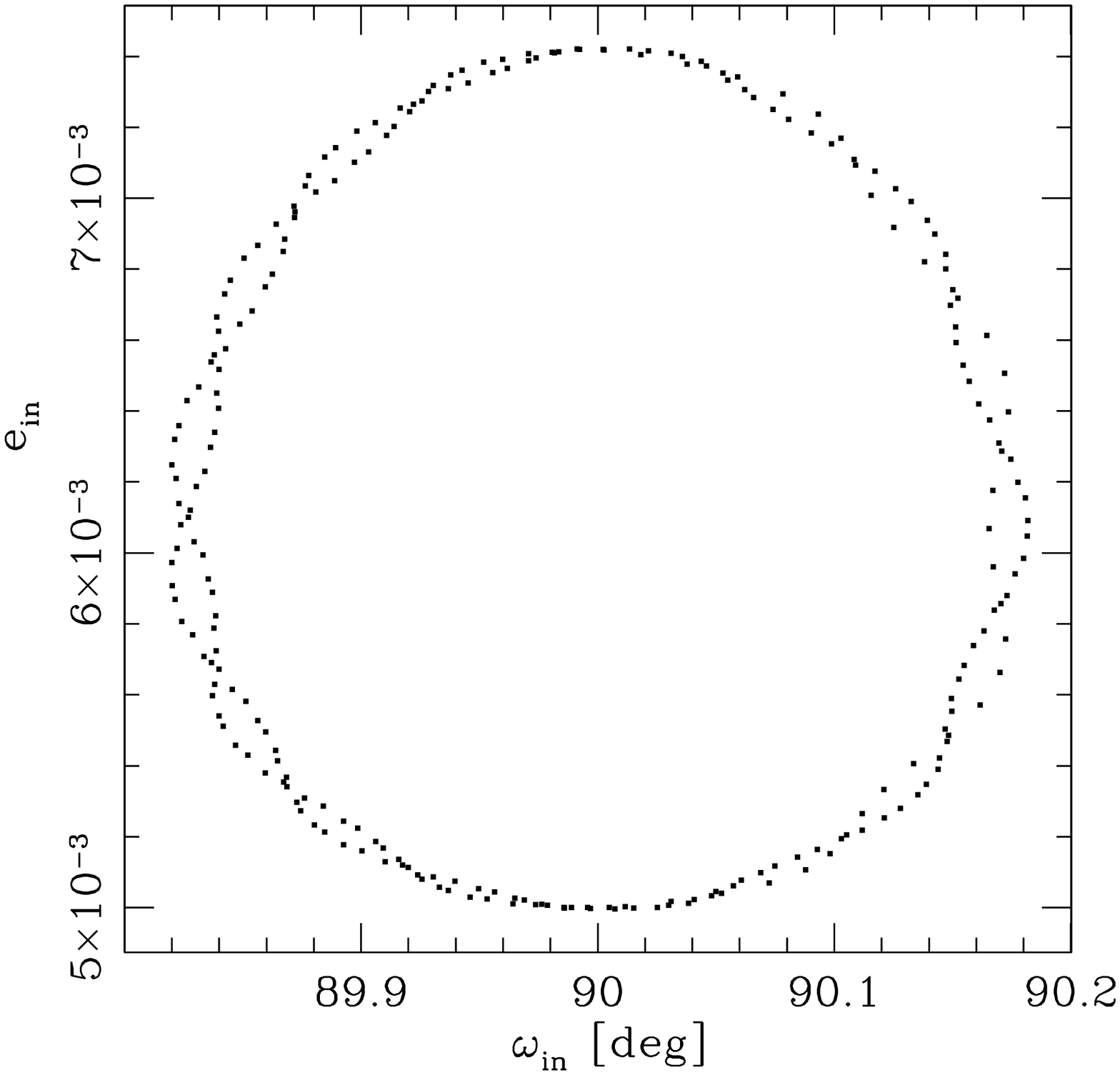}
\caption{M15 X-2: Same as figure \ref{Fig:4U18e}. The long period -- i.e. period of the eccentricity oscillations -- is $3.4\yr$.
 \label{Fig:M15e}}
\end{figure}

Figures \ref{Fig:4U18e}-\ref{Fig:M15e} show the eccentricity oscillations with corresponding long periods for UCXB 4U 1850-087, 4U 0513-40 and M15 X-2. The amplitude of the eccentricity oscillations is $\sim 5\times10^{-3}$ in the case of 4U 1850-087 and 4U 0513-40, while in the case of M15 X-2 the amplitude of eccentricity oscillations is $\sim 3\times10^{-3}$. The observed factor $\sim 2-3$ luminosity variations are explained by these eccentricity oscillations  \citep{2007MNRAS.377.1006Z}. Trapping the system in libration around the fixed point provides a natural explanation for  the origin of the long periods and the small eccentricity oscillations that lead to observed luminosity variations. 

In order to keep the fiducial eccentricity of M15 X-2 below the maximum eccentricity given in Table 1 at all times and to have sufficiently large eccentricity oscillation to produce luminosity variations of a factor of $2-3$, we require in our numerical calculations a long period of $3.4\yr$, which is a factor of $\sim 3$ larger than the estimated minimum long period listed  in Table 1. Note that any long period variations in luminosity for M15 X-2 remain undetected until this date.


\begin{table*}
\begin{center}
\begin{tabular}{|l|l|l|l|}
\tableline
\multicolumn{4}{|c|}{TABLE 3. 4U 0513-40: System parameters} \\
\tableline
\tableline
Symbol & Definition & Value & Citation \\ \hline
$m_1$ & Neutron star (primary) mass & $1.4 M_{\odot}$ &\\
$m_2$ & White dwarf (secondary) mass & $0.045 M_{\odot}$ & see section \ref{sec:masses}\\
$m_3$ & Third companion mass & $0.55 M_{\odot}$& \\
$a_1$ & Inner binary semimajor axis & $1.65\times10^{10}\cm$ & \citet{2009ApJ...699.1113Z}\\
$a_{out}$ & Outer binary semimajor axis & $8.36a_1$ &\\
$e_{in, 0}$ & Inner binary initial eccentricity & $0.02$ &\\
$e_{out, 0}$ & Outer binary eccentricity & $10^{-4}$ &\\
$i_{init}$ & Initial mutual inclination & $46.377^{o}$  & \\
$\omega_{in, 0}$ & Initial argumet of periastron & $90^{o}$ &\\
$\Omega_{in}$ & Longitude of ascending node & $0$ &\\
$R_2$ & White dwarf radius & $2.64\times10^9 \cm$ &see section \ref{sec:masses} \\
$k_2$ & Tidal Love number & $0.01$ & Arras (private communication)\\
$Q$ & Tidal dissipation factor & $5\times10^7$ &\\
\tableline
\tableline
\end{tabular}
\end{center}
\end{table*}



\begin{table*}
\begin{center}
\begin{tabular}{|l|l|l|l|}
\tableline
\multicolumn{4}{|c|}{TABLE 4. M15 X-2 : System parameters} \\
\tableline
\tableline
Symbol & Definition & Value & Citation \\ \hline
$m_1$ & Neutron star (primary) mass & $1.4 M_{\odot}$ &\\
$m_2$ & White dwarf (secondary) mass & $0.034 M_{\odot}$ & see section \ref{sec:masses}\\
$m_3$ & Third companion mass & $0.55 M_{\odot}$& \\
$a_1$ & Inner binary semimajor axis & $2.1\times10^{10}\cm$ & \citet{2005ApJ...634L.105D}\\
$a_{out}$ & Outer binary semimajor axis & $9.5a_1$ &\\
$e_{in, 0}$ & Inner binary initial eccentricity & $0.015$ &\\
$e_{out, 0}$ & Outer binary eccentricity & $10^{-4}$ &\\
$i_{init}$ & Initial mutual inclination & $44.643^{o}$  & \\
$\omega_{in, 0}$ & Initial argumet of periastron & $90^{o}$ &\\
$\Omega_{in}$ & Longitude of ascending node & $0$ &\\
$R_2$ & White dwarf radius & $2.75\times10^9 \cm$ & see section \ref{sec:masses}\\
$k_2$ & Tidal Love number & $0.01$ & Arras (private communication)\\
$Q$ & Tidal dissipation factor & $6\times10^7$ &\\
\tableline
\tableline
\end{tabular}
\end{center}
\end{table*}


\subsection{Resonant trapping and detrapping}

As argued in \citet{2012ApJ...747....4P} the semimajor axis of the inner binary has to expand. This expansion is expected in the standard evolutionary scenario (see Introduction). As the orbit expands due to the mass transfer,  the action of the separatrix increases adiabatically on the accretion timescale, leading to trapping in a resonance around the fixed point. One might expect that tidal effects could be dominant at such a small separation. There are two arguments against a shrinking semimajor axis. The first argument is that the phase where the semimajor axis shrinks and the eccentricity decays due to tidal dissipation is short-lived; only a few thousand years until the mass transfer driven expansion  dominates the evolution. The second argument is that the shrinking orbit drives the initially librating orbit out of resonance into circulation which would change dramatically the period of luminosity variations.  Thus, as shown previously for the case of 4U 1820-30, the evolution of the system is dominated by mass transfer but the rate of expansion of the orbit is  decreased due to tidal dissipation. 

\begin{figure}
\epsscale{0.7} 
\plotone{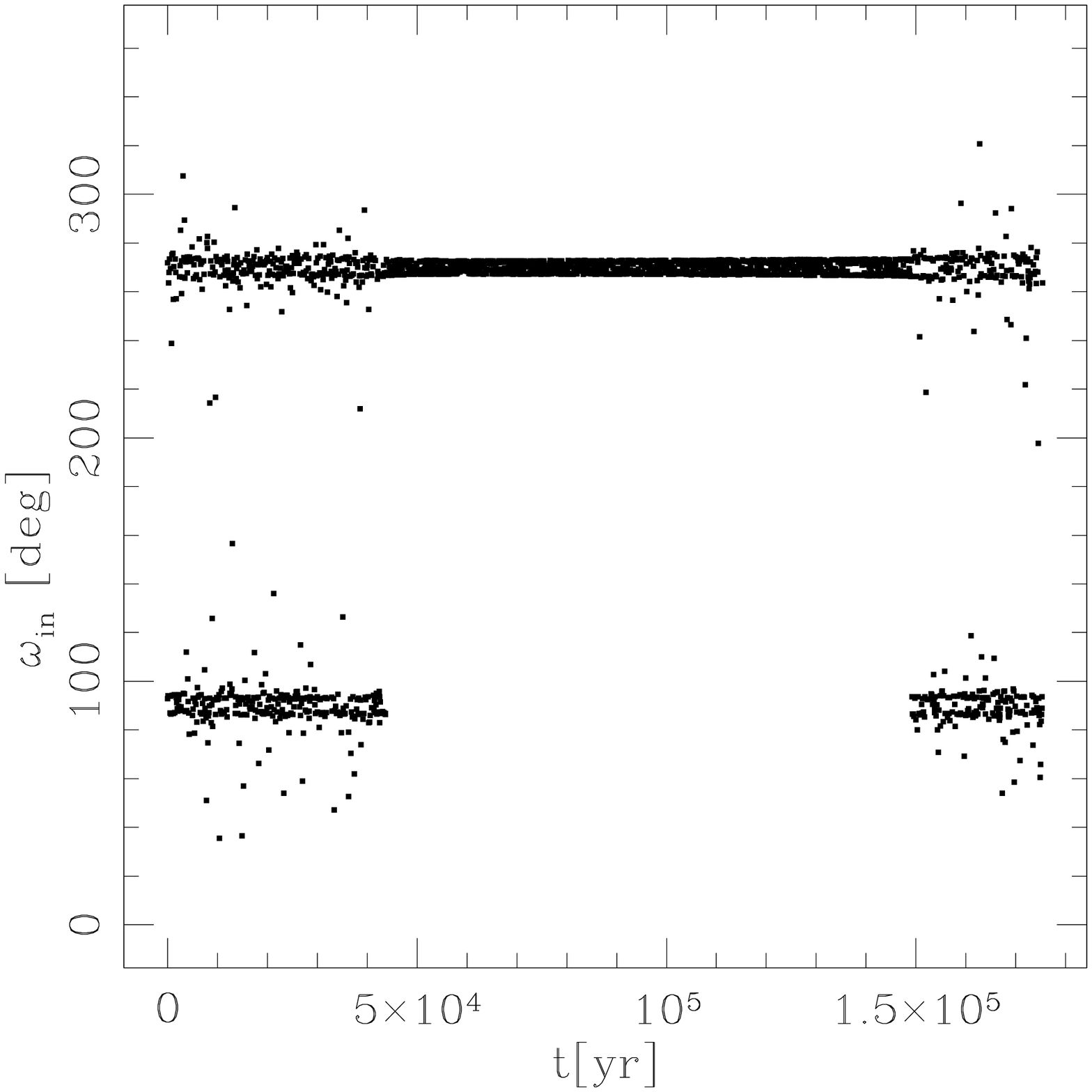}
\plotone{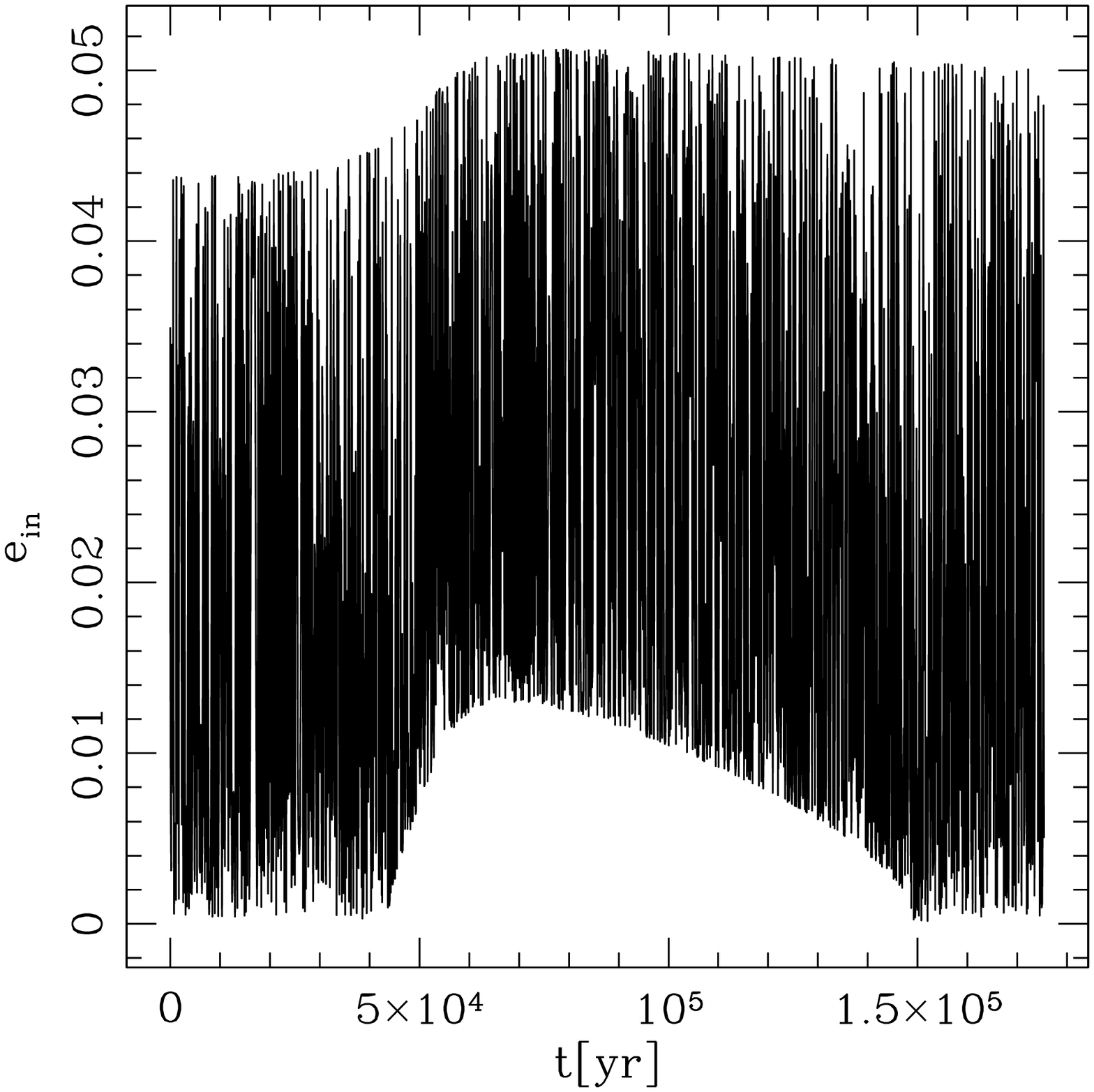}
\caption{4U 1850-087: a) $\omega$ vs $t$. Initially we place the system on
  a circulating orbit. The system is trapped in libration after $\approx 40000\yr$. We set $Q=1\times10^8$ and the initial eccentricity i $e_0=0.044$; all remaining parameters have values listed in Table 2. For this choice of parameters, the system remains in libration for about $3\times10^5\yr$. b)The eccentricity as a function of time -- does not exceed significantly the estimated maximum value of $0.05$.  \label{Fig:4U18_trap}}
\end{figure}

\begin{figure}
\epsscale{0.7} 
\plotone{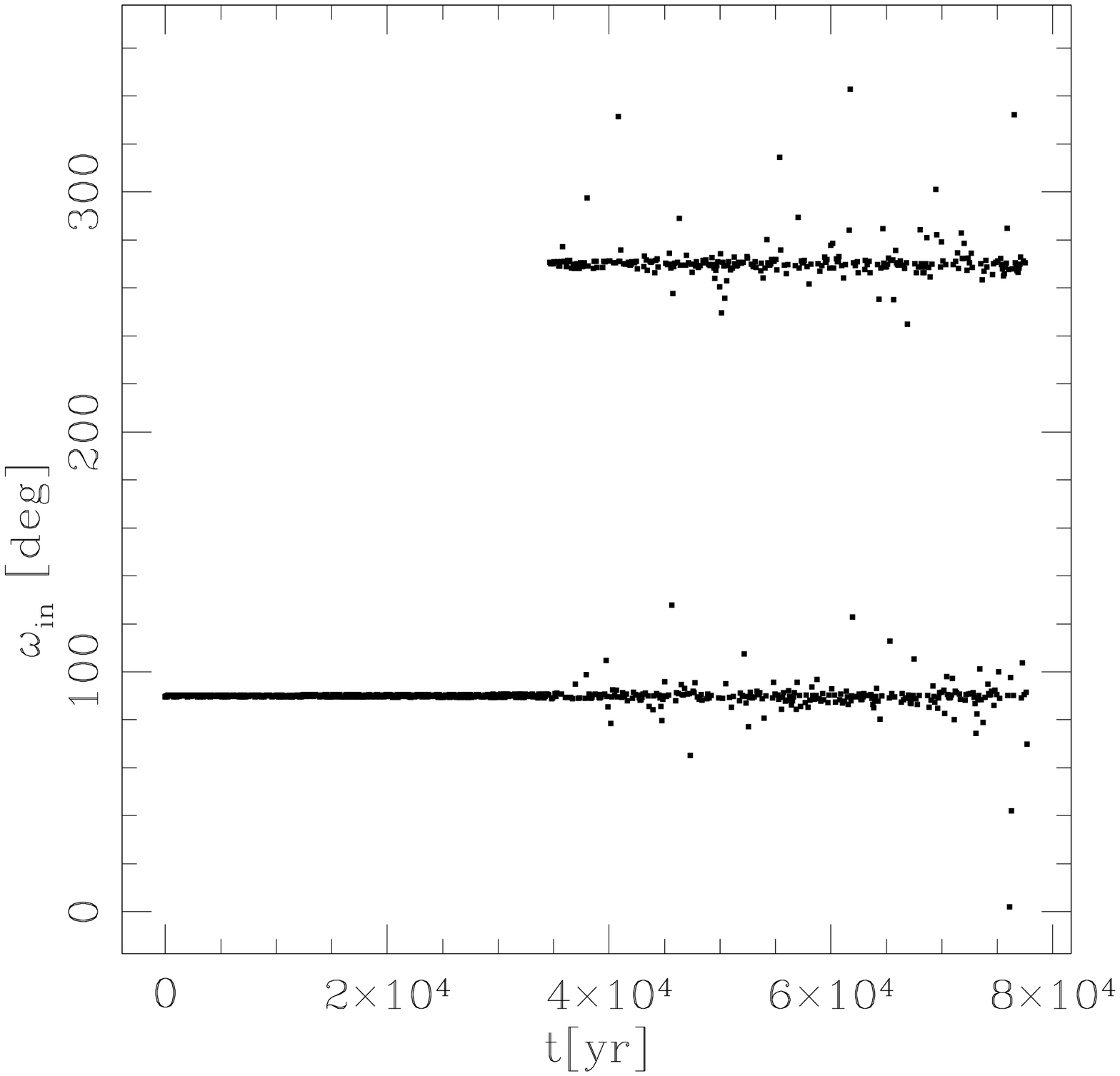}
\plotone{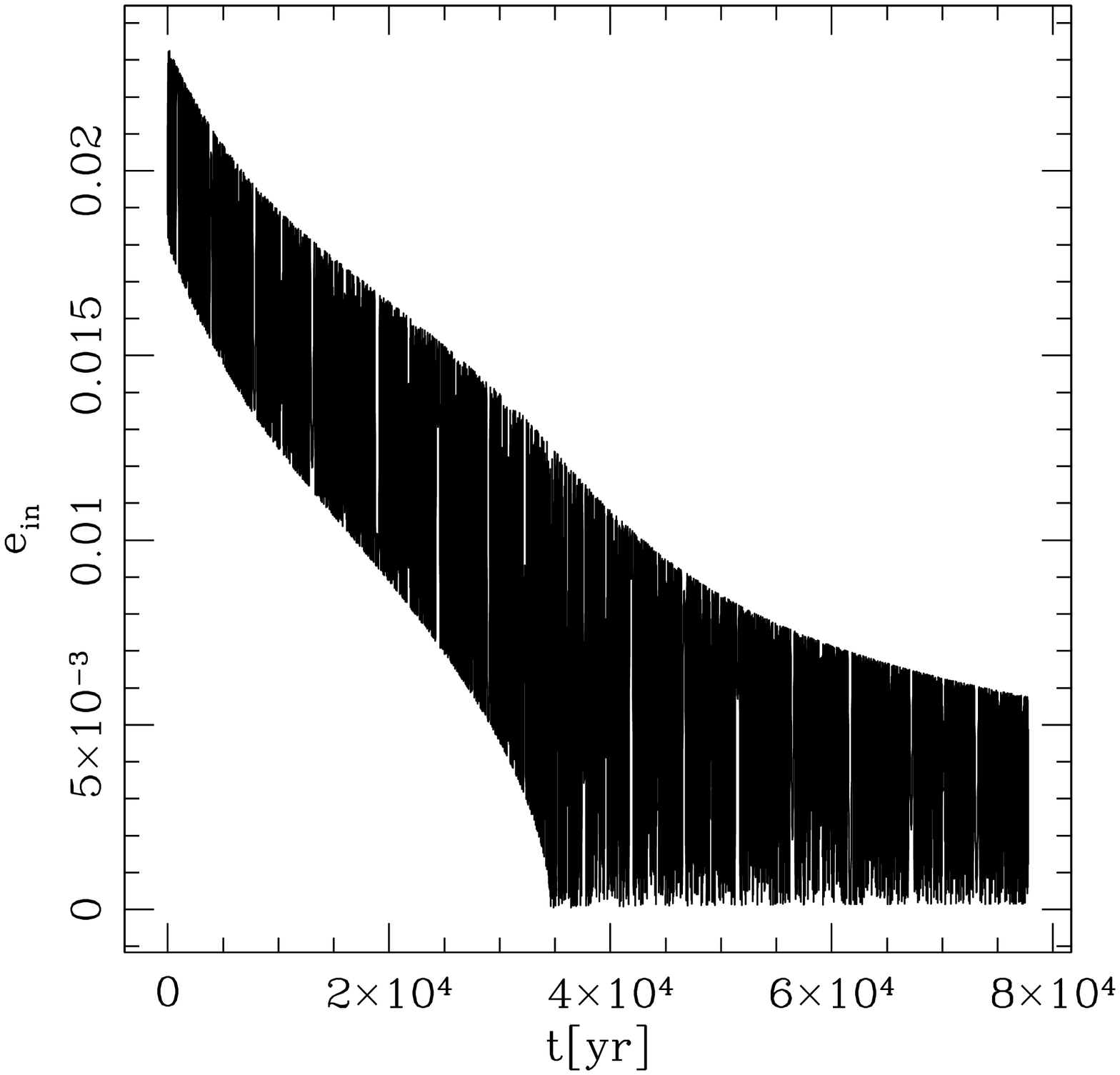}
\caption{4U 1850-087: a)  $\omega$ vs $t$. Initially the system is placed on librating orbit. We set $Q=2\times10^{7}$, which leads to shrinking of the semimajor axis and the
action of the separatrix is decreasing. Consequently the system is driven out from
the resonance after $\approx 35000\yr$. b) The eccentricity as a function of time.
 \label{Fig:4U18_detrap}}
\end{figure}

\begin{figure}
\epsscale{0.7} 
\plotone{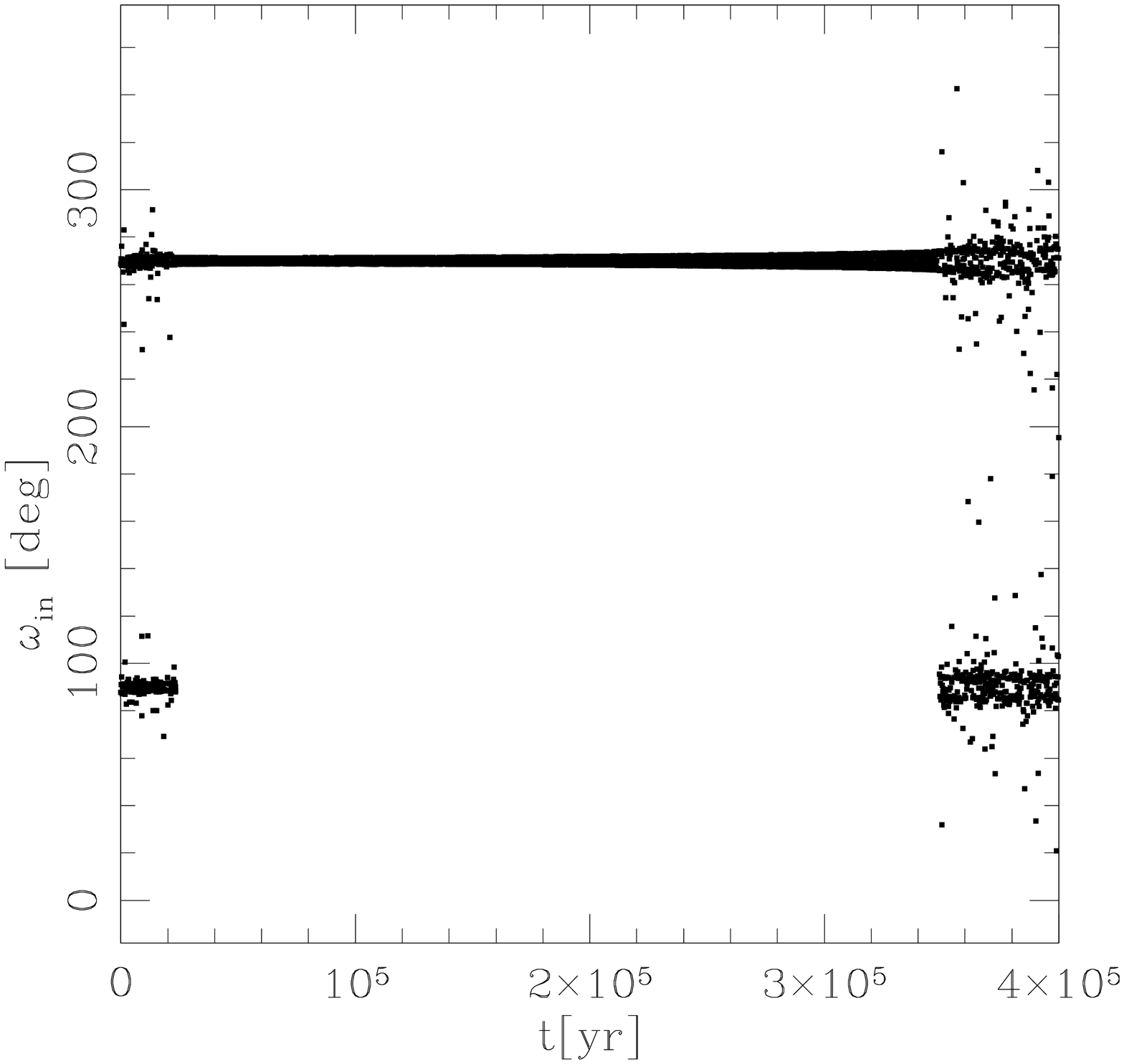}
\plotone{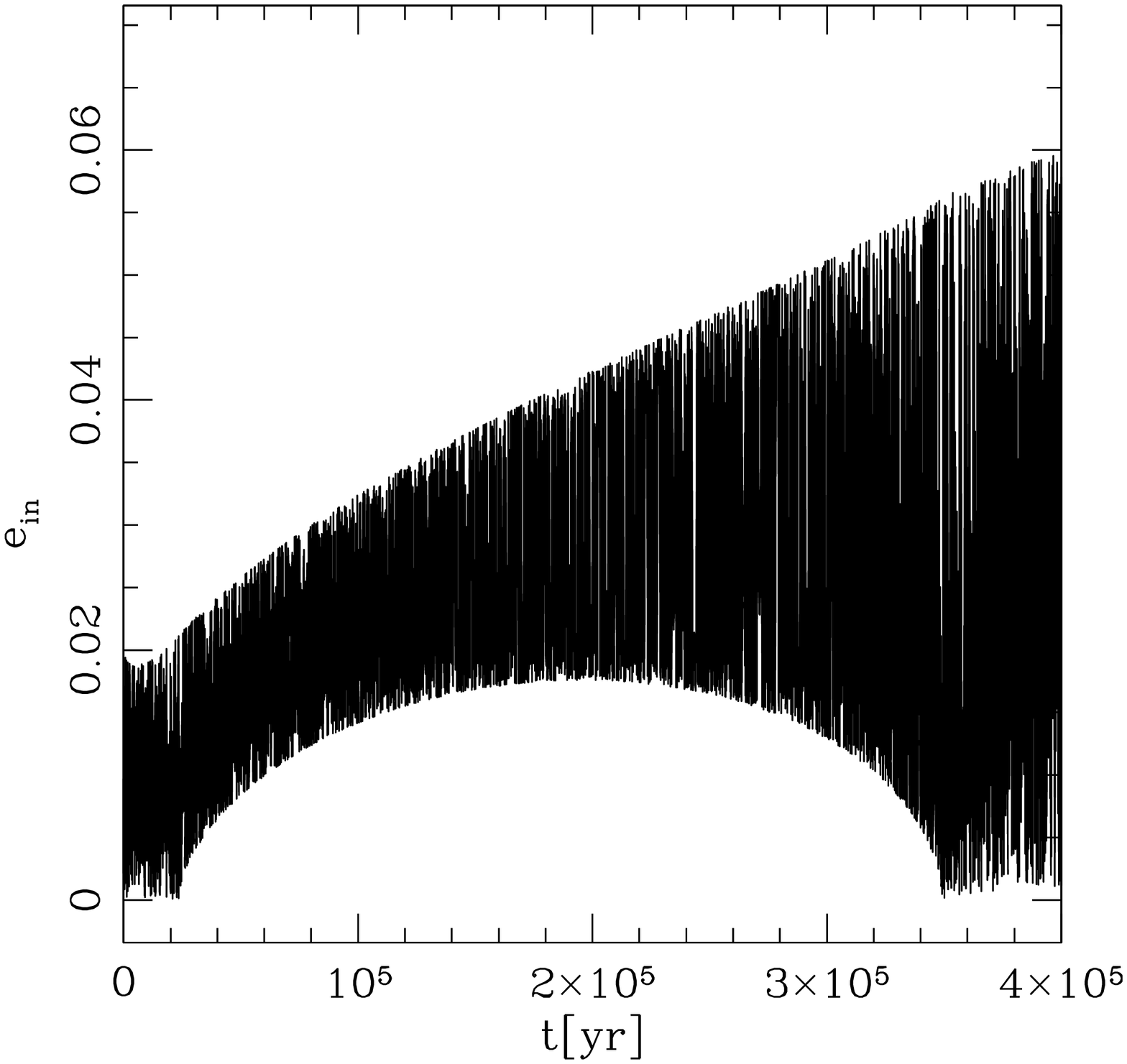}
\caption{4U 0513-40: Same as Figure \ref{Fig:4U18_trap} with $Q=5\times10^7$ and $e_0=0.02$. Trapping in libration occurs after $\approx 2\times 10^4\yr$. The eccentricity does not exceed significantly the estimated maximum value of $0.05$ for at least $3\times10^5\yr$, a time it spends trapped in libration.  \label{Fig:4U05_trap}}
\end{figure}

\begin{figure}
\epsscale{0.7} 
\plotone{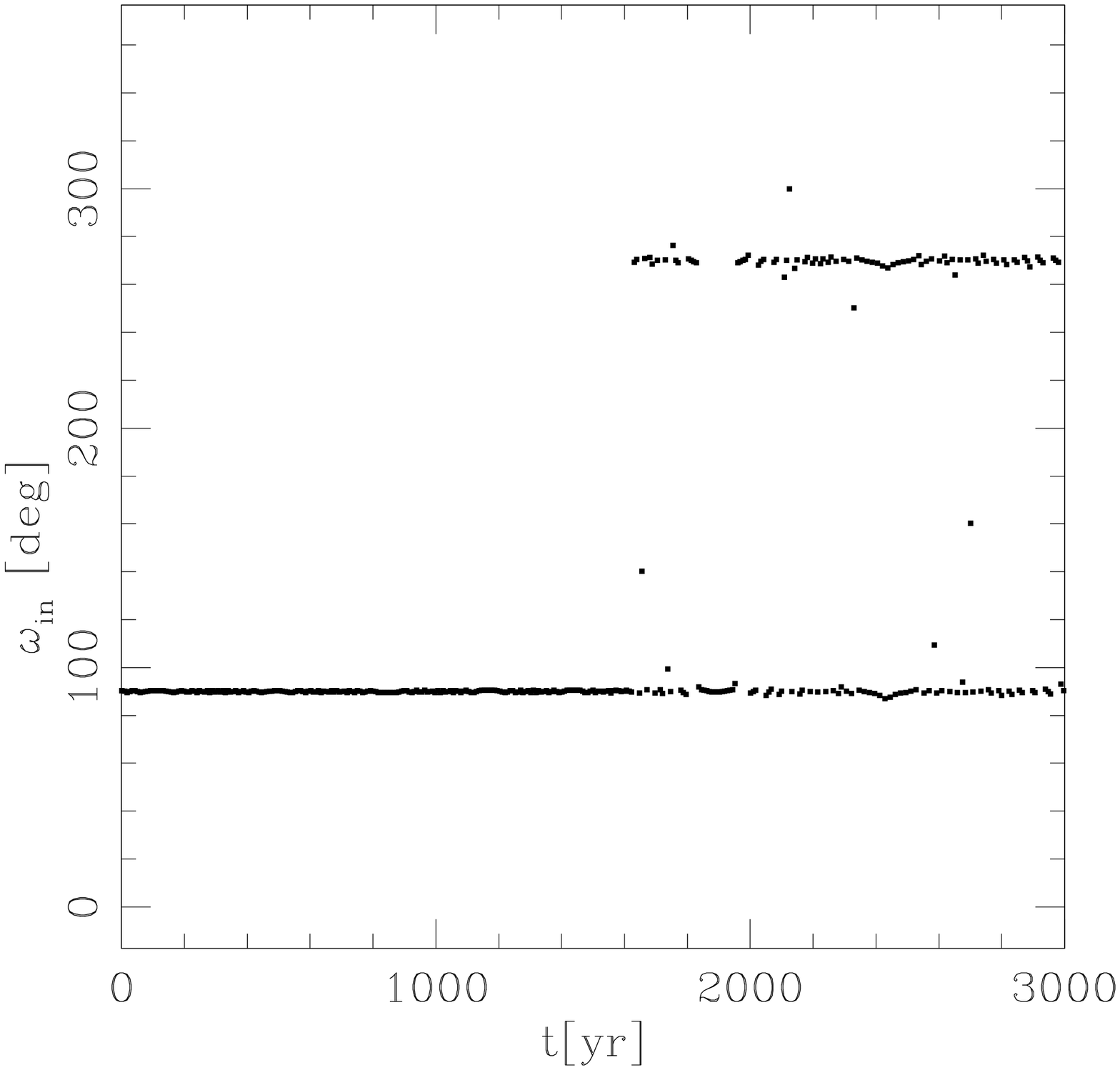}
\plotone{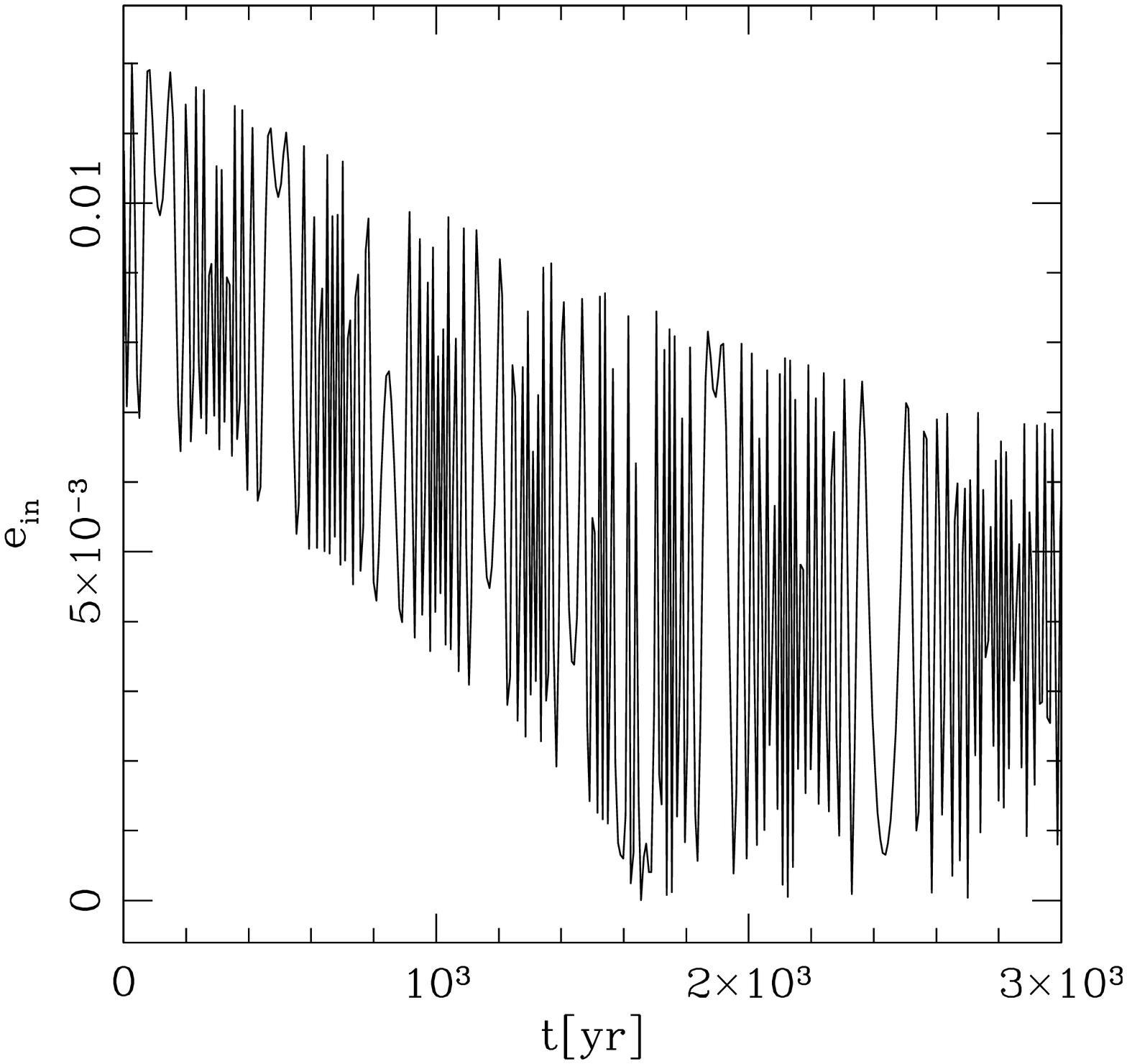}
\caption{4U 0513-40: Same as Figure \ref{Fig:4U18_detrap}. Resonance detrapping occurs after $\approx 1800\yr$. \label{Fig:4U05_detrap}}
\end{figure}

\begin{figure}
\epsscale{0.7} 
\plotone{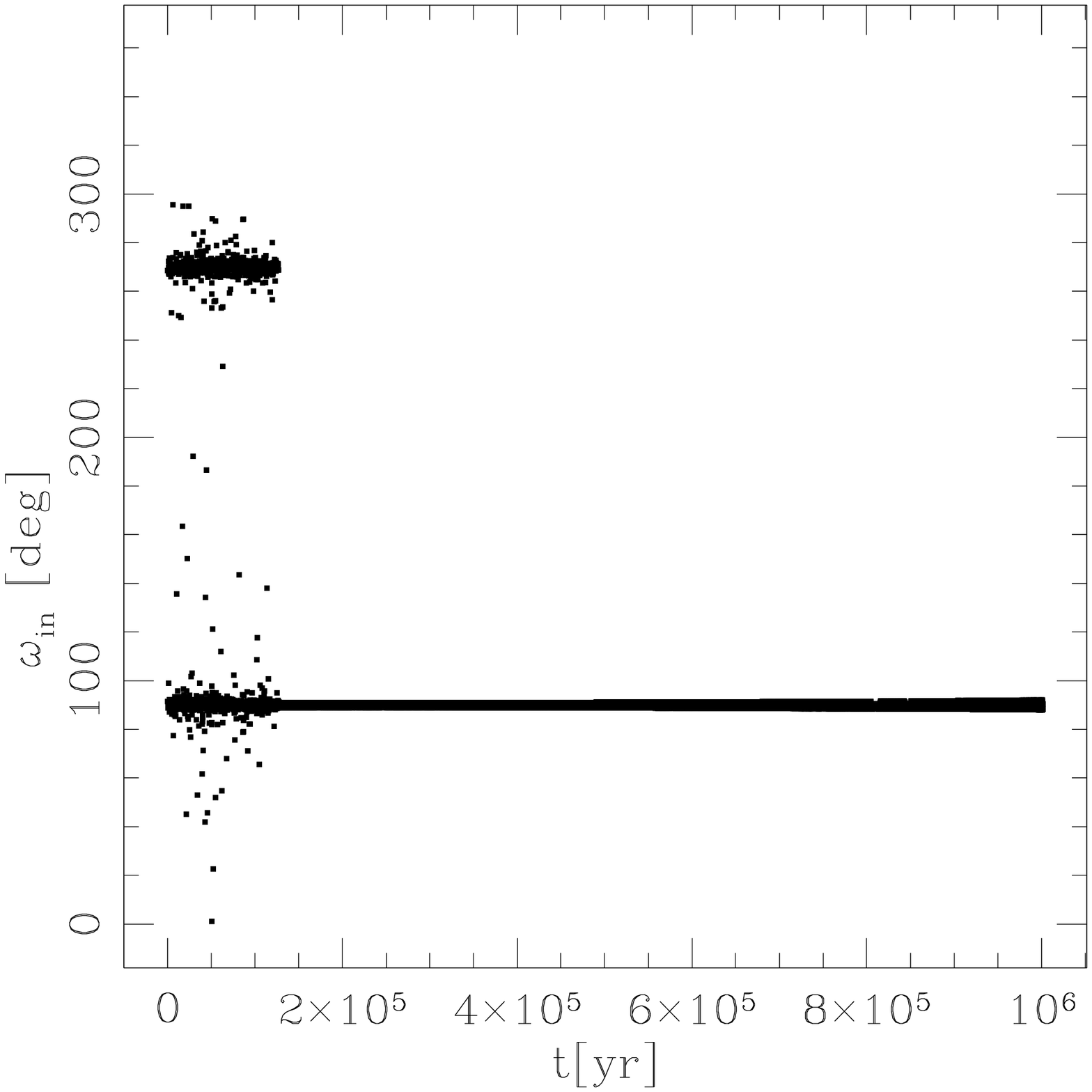}
\plotone{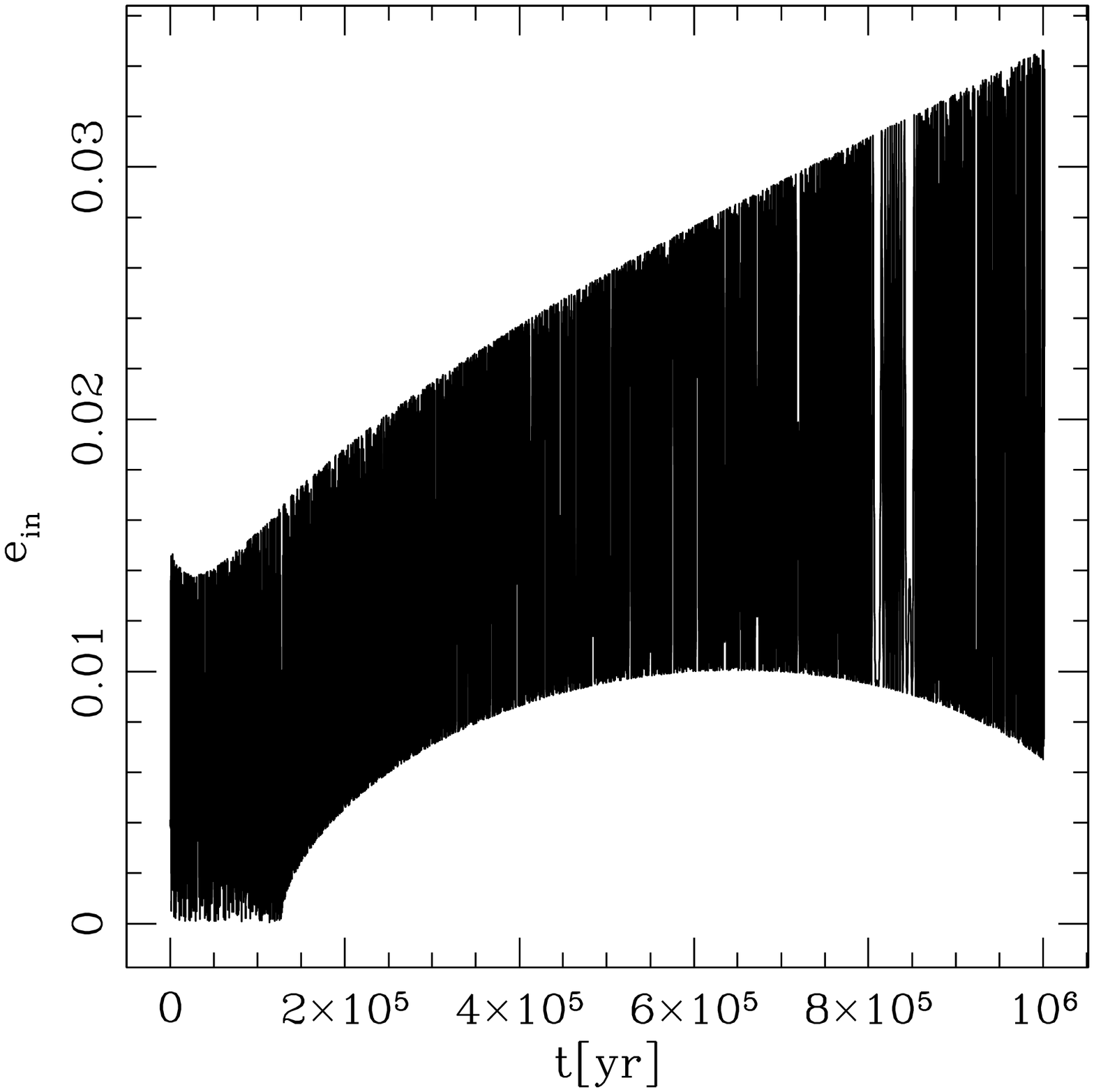}
\caption{M15 X-2: Same as Figure \ref{Fig:4U18_trap} with $Q=6\times10^7$ and $e_0=0.015$. After $\approx 20000\yr$ the system gets captured in libration where it remains for about  $3\times10^5\yr$. During that time the eccentricity does not exceed significantly the estimated maximum value of $0.04$. 
 \label{Fig:M15_trap}}
\end{figure}

\begin{figure}
\epsscale{0.7} 
\plotone{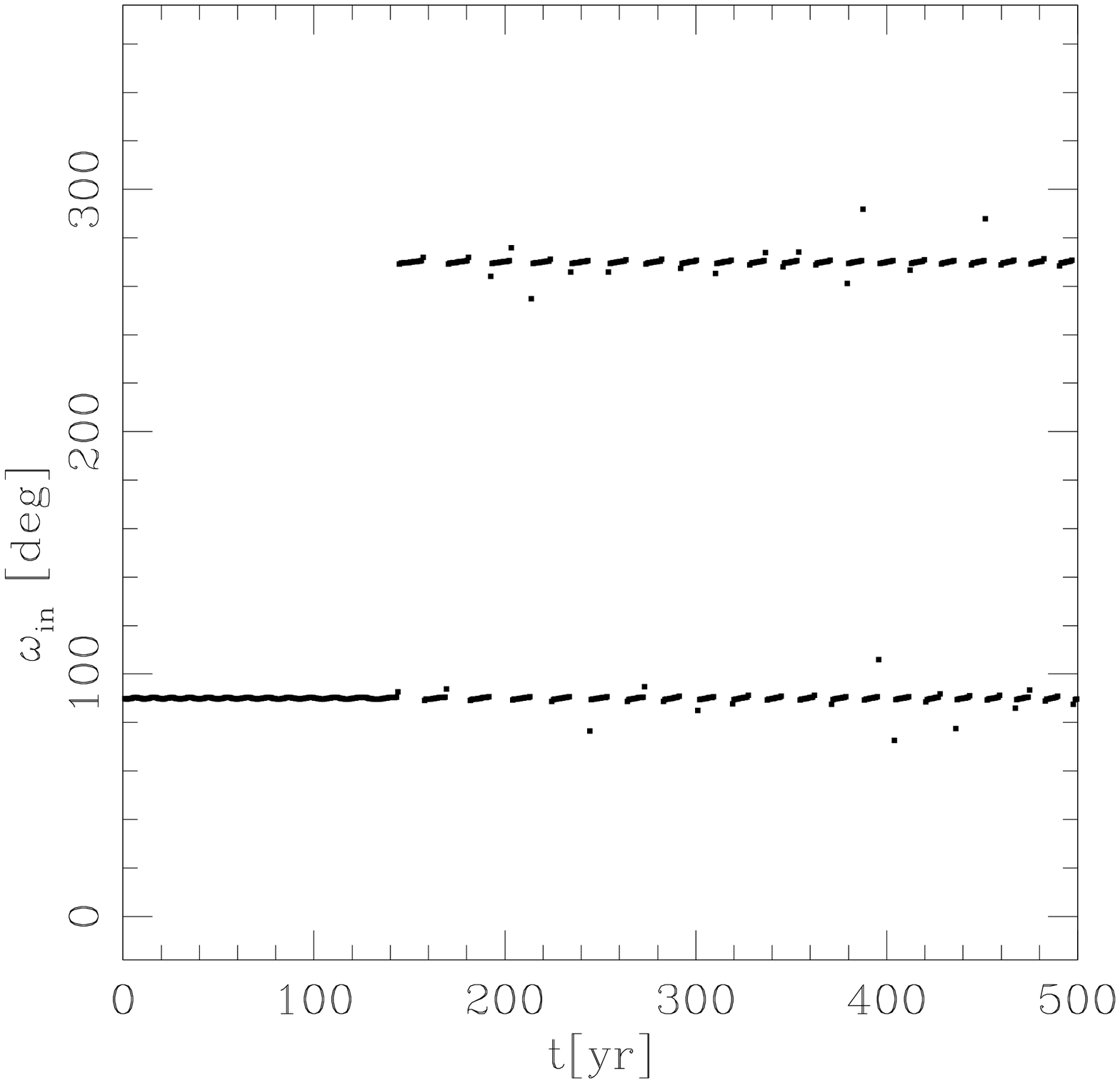}
\plotone{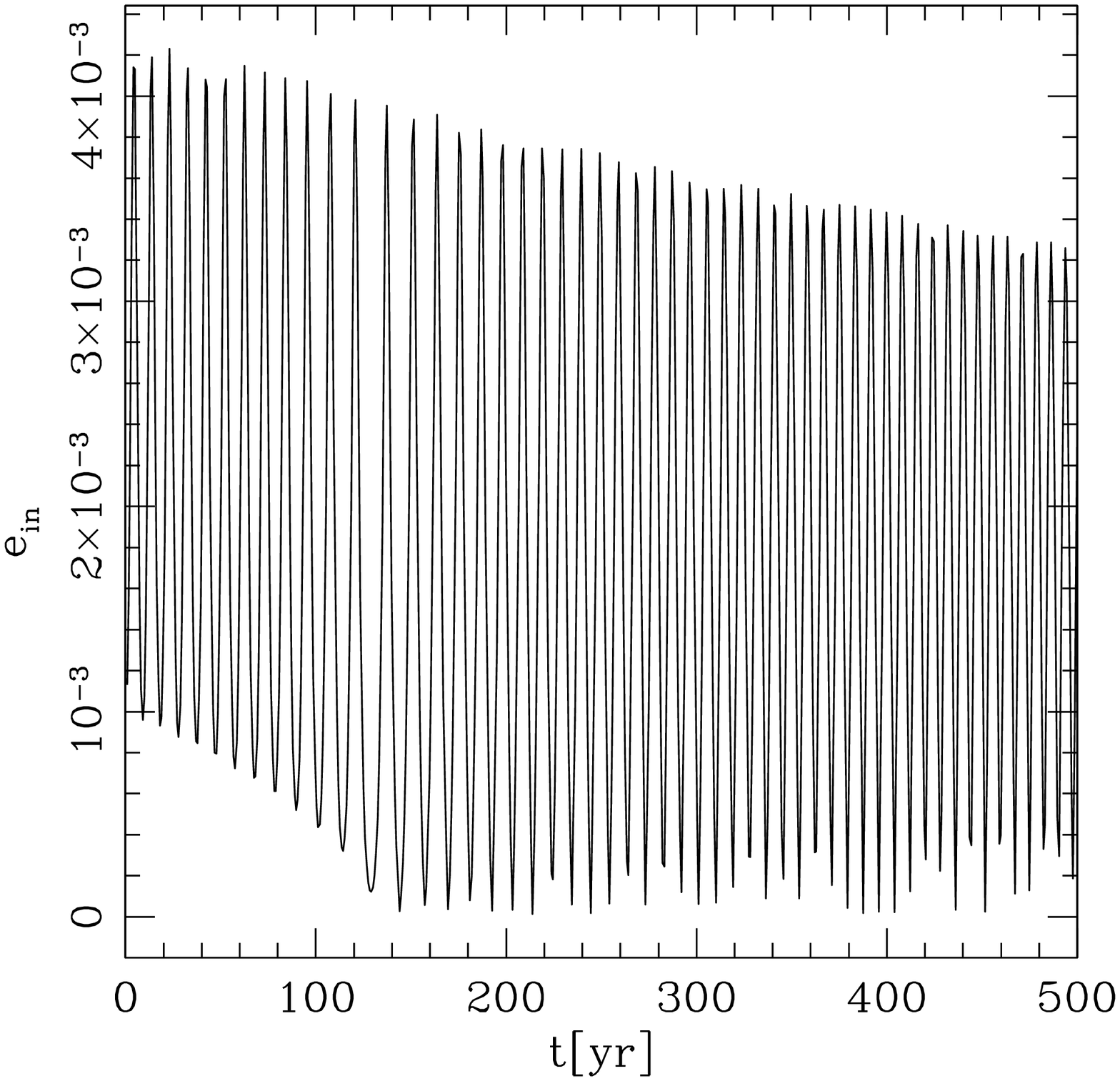}
\caption{M15 X-2: Same as Figure \ref{Fig:4U18_detrap}. Resonance detrapping occurs after $\approx 140\yr$.\label{Fig:M15_detrap}}
\end{figure}

In this section we demonstrate that just this physical picture applies in the case of these three binaries as well and it can indeed explain the origin of their long periods. Figures \ref{Fig:4U18_trap}, \ref{Fig:4U05_trap} and \ref{Fig:M15_trap} show the systems initially put on circular orbit with a choice of $Q$ such that the orbit expands. As integration proceeds, the separatrix continues to expand and  at some point captures the initially circulating orbit in libration around the fixed point. Once captured, the system librates for at least $10^5\yr$.  We consider $10^5\yr$ to be a reasonable fraction of the system's lifetime during which it can be observed in such a state (see section \ref{sec:Qconstraint}). Figures \ref{Fig:4U18_detrap}, \ref{Fig:4U05_detrap} and \ref{Fig:M15_detrap}, on the other hand, show the case where the systems is initially on librating orbit with $Q$ such that semimajor axis shrinks. In all three cases detrapping from the resonance occurs fairly quickly making observation of the system in such a state highly unlikely.

\section{CONSTRAINING THE TIDAL DISSIPATION FACTOR Q FOR THE WHITE DWARF COMPANIONS}\label{sec:Qconstraint}

\begin{figure}
\epsscale{0.7} 
\plotone{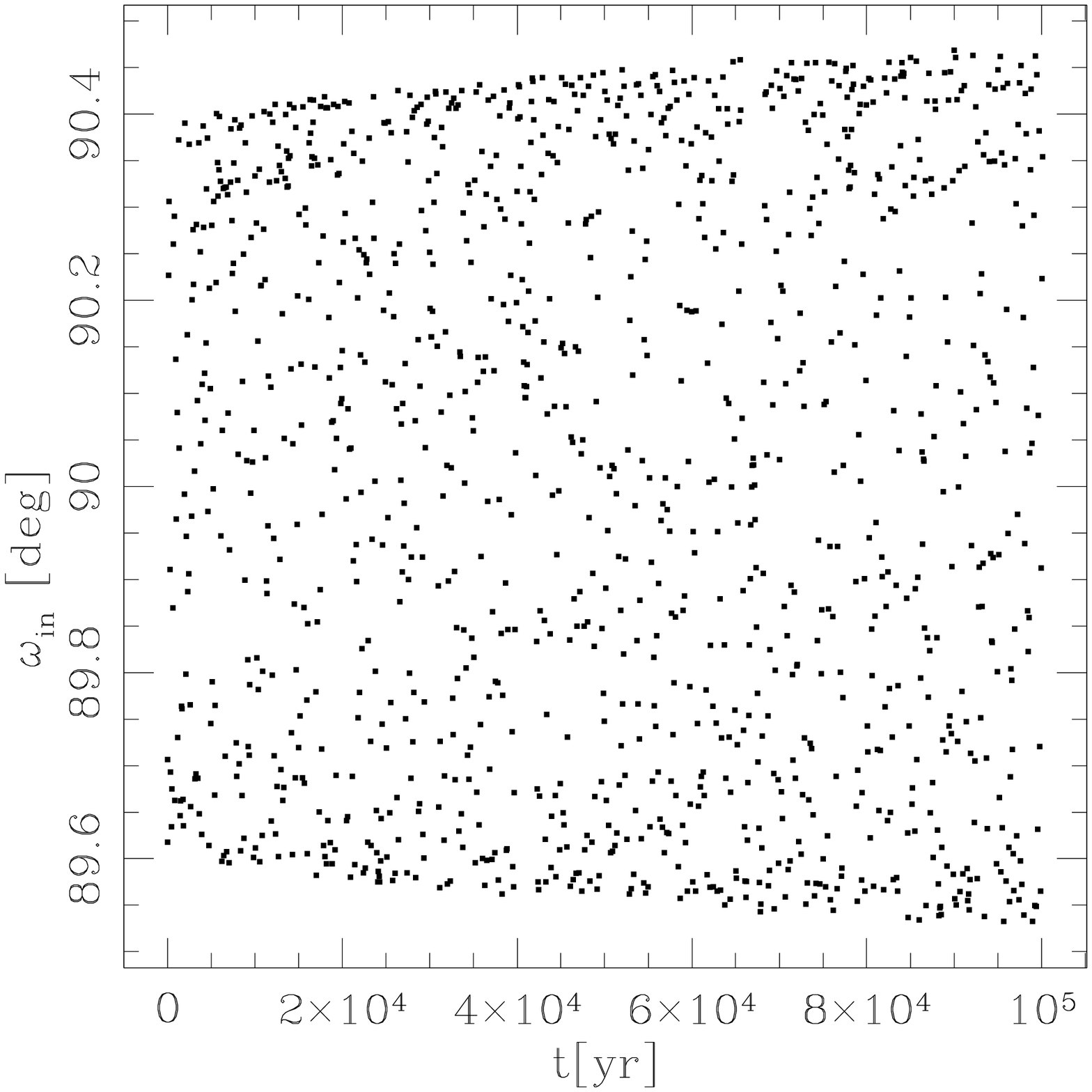}
\plotone{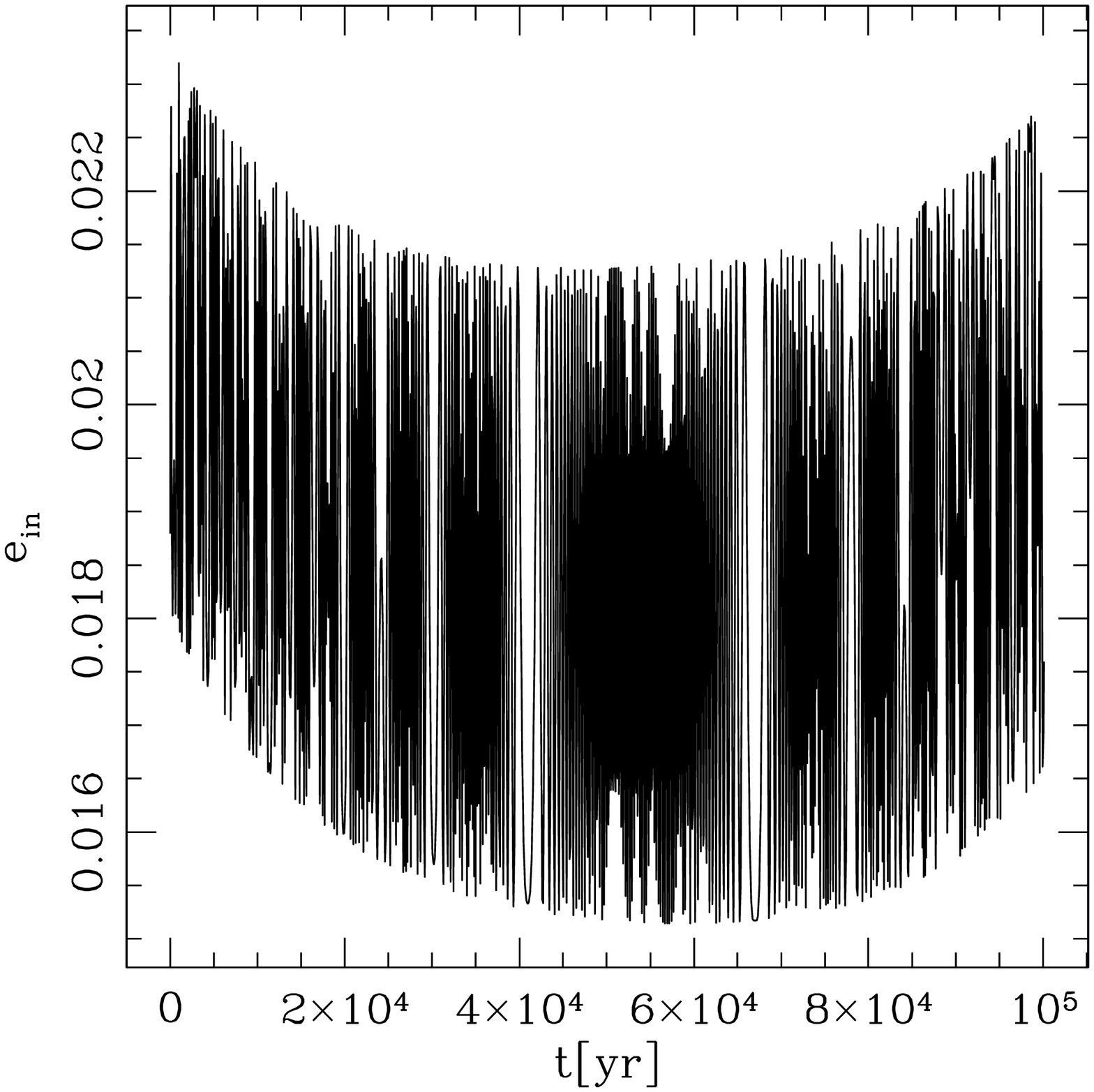}
\caption{4U 1850-087: a) $\omega$ vs $t$. Here we have used $Q=6\times10^7$; remaining parameters are as listed in Table 2. For this choice of parameters, the system remains in libration for about $10^5\yr$. b) The eccentricity as a function of time. The eccentricity does not exceed the estimated maximum value of $0.05$ during the integration.  \label{Fig:4U18_stay}}
\end{figure}

\begin{figure}
\epsscale{0.7} 
\plotone{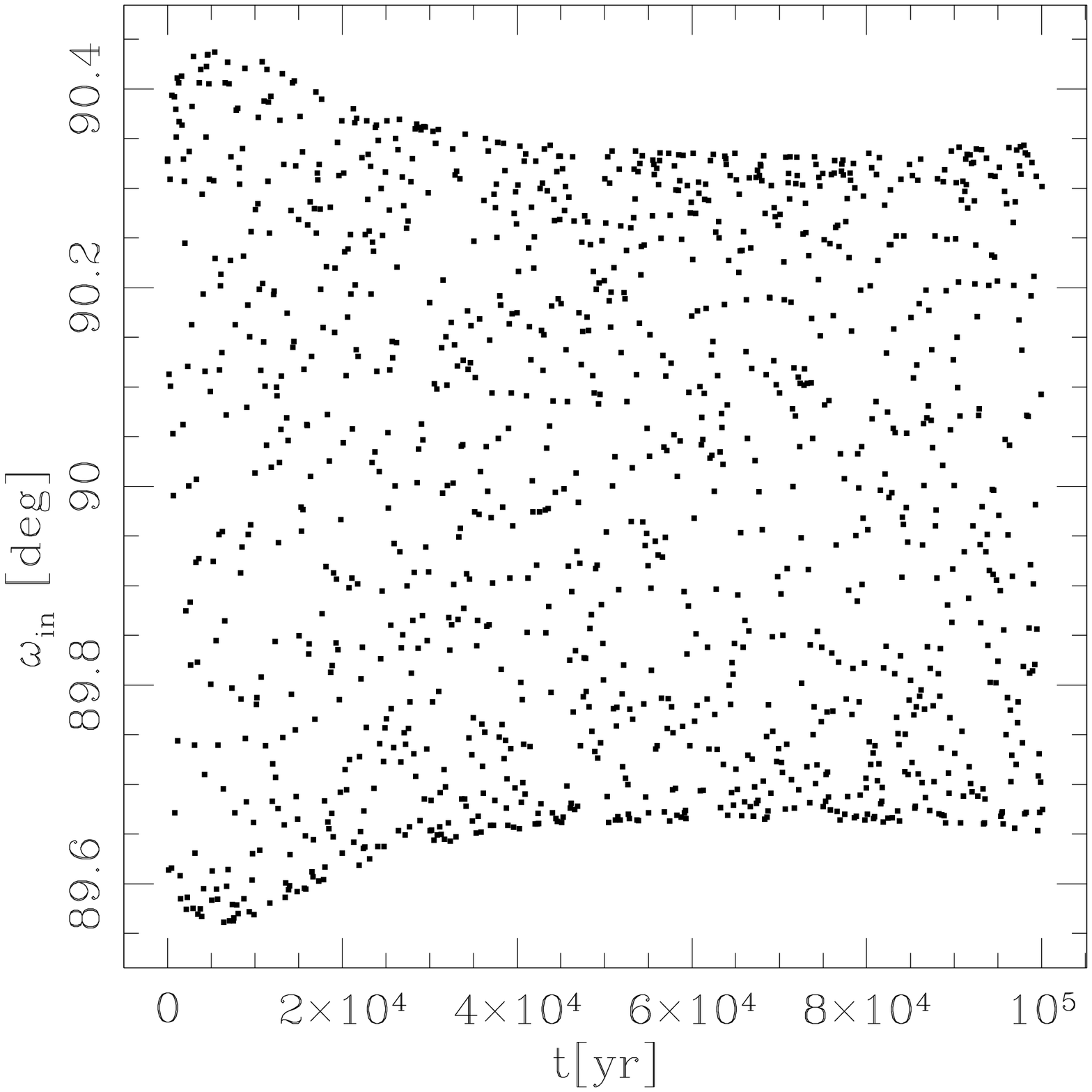}
\plotone{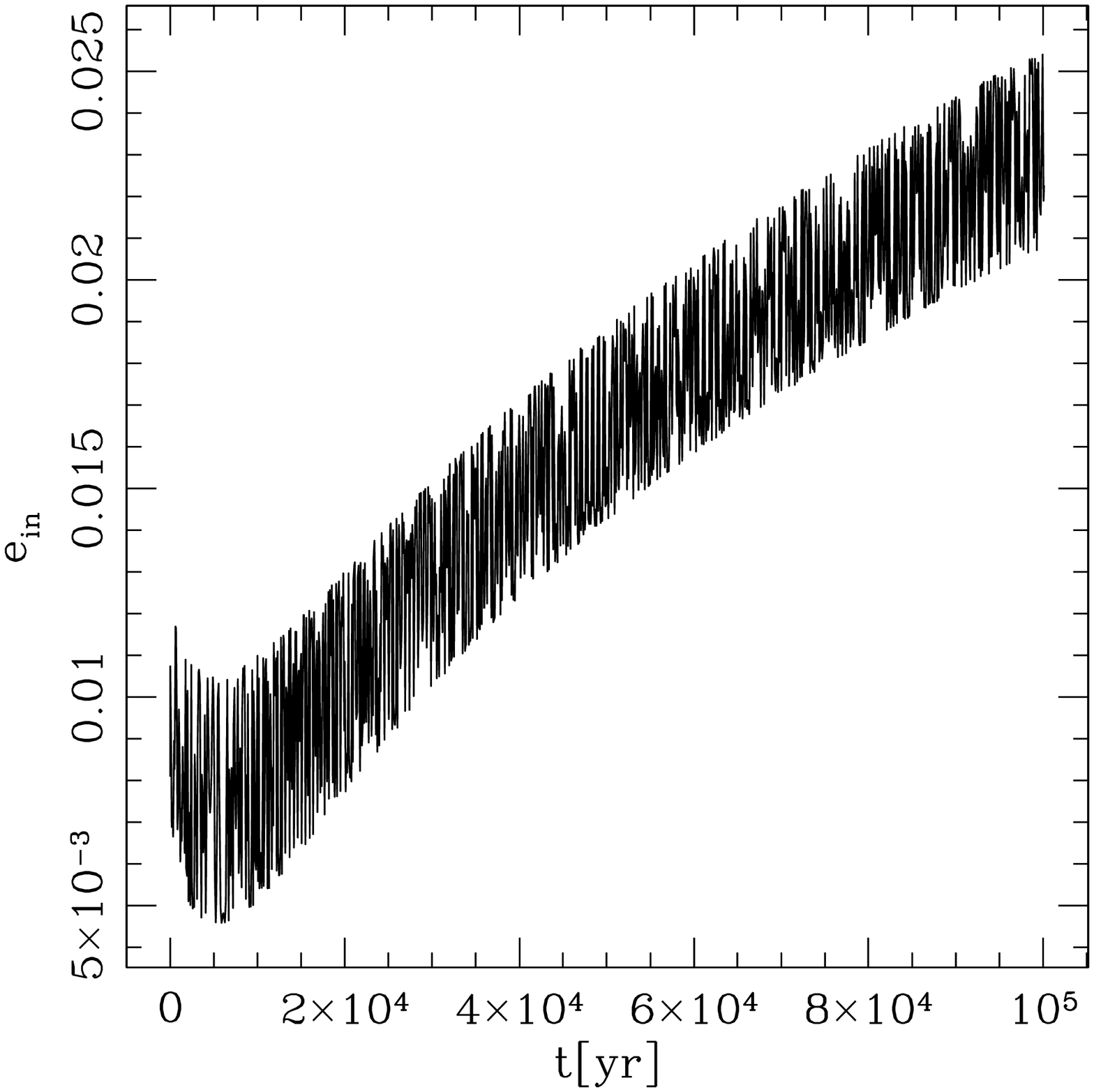}
\caption{4U 0513-40: Same as Figure \ref{Fig:4U18_stay} with $Q=5\times10^7$. \label{Fig:4U05_stay}}
\end{figure}

\begin{figure}
\epsscale{0.7} 
\plotone{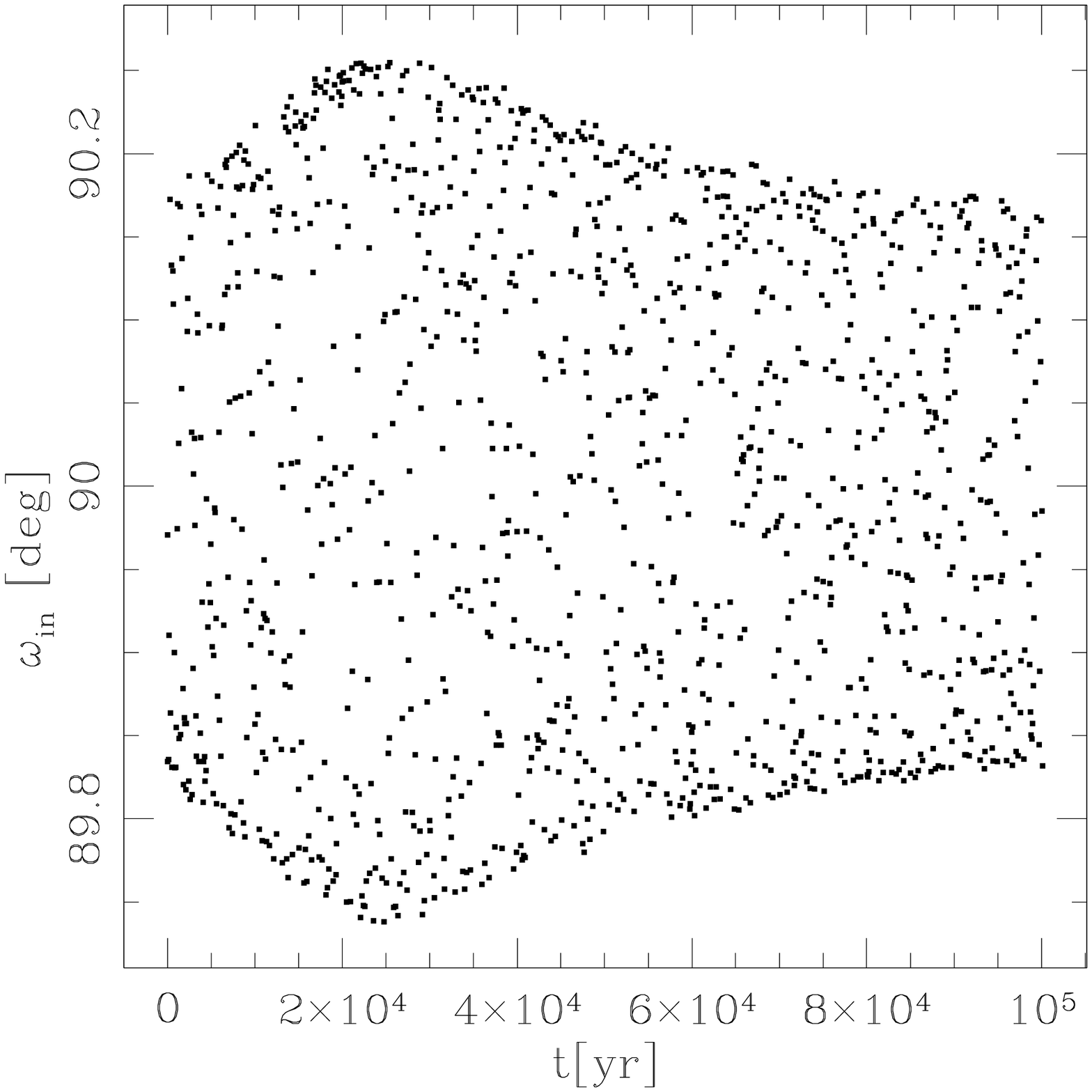}
\plotone{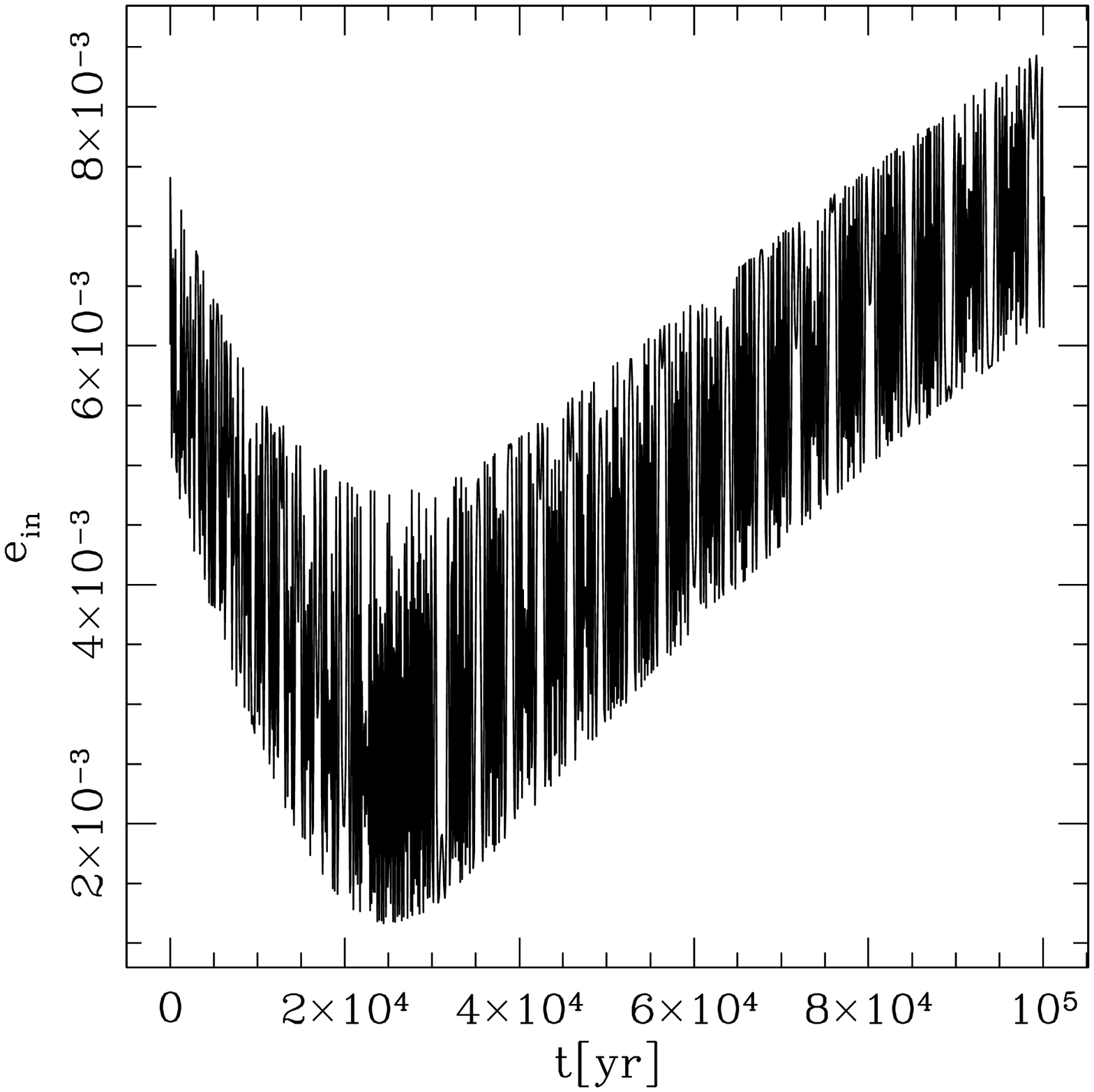}
\caption{M15 X-2: Same as Figure \ref{Fig:4U18_stay} with $Q=6\times10^7$.   \label{Fig:M15_stay}}
\end{figure}

In this section we constrain the tidal dissipation factor $Q$ for the white dwarf companions simply by asking that the system stays in the resonance for more than $10^5\yr$. Estimated mass transfer rates for these systems are $(5\times10^{-10}$-$10^{-9})M_{\odot}/\yr$ (see equation \ref{eqn:mdot}) giving the lifetime of $\sim7\times10^7\yr$ during which these systems can sustain the observed luminosity. Therefore a reasonable fraction of time to stay in the resonance is indeed at least $10^5\yr$.  Figures \ref{Fig:4U18_stay} to \ref{Fig:M15_stay} demonstrate that for the fiducial values of $Q$ or more precisely $e^2Q/k_2$ listed in table 5 these systems remain trapped in the resonance for reasonable fraction of their lifetime during which the maximum eccentricity does not exceed values given in Table 1.


\begin{table}
\begin{center}
\begin{tabular}{|l|l|l|l|}
\tableline
\multicolumn{2}{|c|}{TABLE 5. Tidal dissipation factor Q} \\ \hline
\multicolumn{2}{|c|}{4U 1850-087} \\
\tableline
\tableline
Symbol & Value \\ \hline
$(\frac{e}{0.018})^2\frac{Q}{k_2}$ & $6\times10^9$\\ \hline
\tableline
\multicolumn{2}{|c|}{4U 0513-40} \\ \hline
Symbol & Value \\ \hline
$(\frac{e}{0.007})^2\frac{Q}{k_2}$ & $5\times10^9$\\ \hline
\tableline
\multicolumn{2}{|c|}{M15 X-2} \\
\tableline
Symbol & Value \\ \hline
$(\frac{e}{0.005})^2\frac{Q}{k_2}$ & $6\times10^9$\\ \hline
\tableline
\end{tabular}
\end{center}
\end{table}


\section{DISCUSSION}\label{sec:discussion}

A long term luminosity periodicity in UCXBs has been suspected for a while now but the observations are not sufficiently good to confirm it with reasonable certainty. The only system for which the long periodicity is certain is the $11\mins$ binary 4U 1820-30. The detection of X-ray bursts in this object \citep{1976ApJ...205L.127G} led to extensive observations and hence a very well sampled light curve. Several authors suggested that this long periodicity may be caused by a third body \citep{1988IAUS..126..347G, 2001ApJ...563..934C, 2007MNRAS.377.1006Z}. In \citet{2012ApJ...747....4P} we attribute the origin of the long period variations in the luminosity to the libration in Kozai resonance with frequency of small oscillations around the fixed point. In addition to the perturbations from a third body, we consider tidal effects, GR precession and mass transfer driven by gravitational wave radiation. Also we show that trapping in a resonance is a consequence of the expansion of the orbit of the inner binary driven by mass-transfer. The model developed for 4U 1820-30 predicts long periodicities in the light curve of 4U 1850-087 and 4U 0513-40 as well as of M15 X-2. Requiring the systems to remain trapped in a resonance for a reasonable fraction of their lifetime allows us to put the upper limit on the tidal dissipation factor for the white dwarf donors. The actual values are listed in Table 5. Obtaining the lower limit for tidal dissipation factor Q of order of $few\times 10^7$ is in agreement with the results of \citet{2011ApJ...740L..53P}, \citet{2011MNRAS.412.1331F}, \citet{2012ApJ...747....4P}.

The three binaries examined in this chapter have similar properties to 4U 1820-30. For two of them, 4U 0513-40 and 4U 1850-087, a long period variations have been suggested \citep{1984ApJ...280..661P, 2010MNRAS.406.2087M}. Therefore we suggest that these three systems may be triples as well and further more we would anticipate that majority of UCXBs in the globular clusters are actually triples. Short-period binaries like these are quite likely to acquire a third body in the dense environment of globular clusters. Comparing the confirmed orbital periods in the field to those in globular clusters, the trend seems to indicate that field UCXBs have orbital periods of order of $40\mins$ while those in globular clusters have periods $\lesssim 20 \mins$.  Such a trend hints at different formation scenarios operating in these two environments. Very long period variations which cannot be due to accretion disk precession or a change in the viewing angle, seem to be the characteristic of UCXBs in globular clusters. To check this speculation it is necessary to determine orbital periods for more of these systems and obtain a better statistical sample.

Several explanations to why the accretion rates differ from those expected based on the binary parameters has been suggested. Other than the presence of the third body, which we discussed in great detail, two viable candidates two are:  tidal disk instabilities \citep{1995PASJ...47L..11O} and irradiation of the donor responsible for the modulation of the mass transfer rate \citep{1986A&A...162...71H}. Their common characteristic is that both cause more stochastic variations than those expected from the presence of a third body. The light curves of these binaries are not sampled as well as the light curve of 4U 1820-30. Even though the data may indicate the potential presence of a regular modulation, there are lots of irregularities that may be the consequence of a combination of regular modulation due to the third body and these other mechanisms causing aperiodic variations.  

The basic idea of the tidal disk instability model relies on the assumption that the mass transfer rate is constant and that all outbursts of accretion onto the primary are caused by intrinsic instabilities in the accretion disk. The disk is compact during the minimum luminosity phase of the long period cycle. The thermal instability generates only quasi-periodic episodes of accretion that are  observed as normal outbursts.  In normal outbursts the accreted mass is less than the mass transferred during the quiescent phase.  The reason for this lies in the inefficiency of  tidal removal of angular momentum from the disk. Continuous built up of the mass and the angular momentum of the disk forces the disk to expand further and further with each consecutive outburst. The expansion continues until the disk reach critical radius for tidal instability. At this point, the final normal outburst sets off the tidal instability creating a precessing eccentric disk. Such precession eccentric disk is observed as a superhump. It enhances greatly the tidal torque, resulting in the superoutburst that  significantly clears out the disk mass. At the end of the described superoutburst, the disk goes back to the initial compact state.

 The second model considers a mass loss instability in the donor star as a consequence of illumination of its  atmosphere by the X-ray flux from the compact object. During the quiescent phase the donor does not fill in completely its Roche lobe, which leads to lower the accretion rate but still sufficient to heat up the external layers of the donor's atmosphere. As these layers are heated up slowly, by an X-ray flux comparable to the stellar flux at the vicinity of the $L_1$ point, they expand. Ultimately, the heating brings the atmosphere in the unstable regime where matter flows at a high rate through the $L_1$ point. Eventually the shielding by the accretion disk may prevent X-ray flux from reaching the $L_1$ region, which will cause the outburst to cease. By this time the heated layers have been transported to the disk. The outburst stops when the entire disk is accreted onto the compact object. 

Tidal instabilities in the accretion disk may explain the variations in the accretion rate on weeks timescale, such as those seen in 4U 0513-40, but most likely not the long period variations \citep{2010MNRAS.406.2087M}. There are no studies of the irradiation induced mass transfer in the context of white dwarfs and hence it is very difficult to make any conclusive statements.  The observations clearly show that irradiation of the white dwarf donor in 4U 0513-40 is significant \citep{2010MNRAS.406.2087M}, but the same is not observed in 4U 1820-30 which is a brighter system. Provided that one finds a reasonable explanation for this discrepancy, the irradiation induced mass transfer model could be feasible. Unquestionably, to understand the details of the dynamics and evolution of UCXBs more observations are required.



\acknowledgments S. P. is grateful to Fabio Antonini and Marcelo Alvarez for providing useful comments on the manuscript. This research has made use of the SIMBAD database, operated at CDS,
Strasbourg, France, and of NASA's Astrophysics Data System.  The
authors are supported in part by the Canada Research Chair program and
by NSERC of Canada.

%
%
%

\bibliography{ucxb}{}

\end{document}